\documentclass[twocolumn,times,tighten]{aastex62}

\pdfoutput=1
\usepackage{apjfonts}
\usepackage{amsmath}

\newcommand{\qso}        {PG 1543$+$489}
\newcommand{\kms}	{km~s$^{-1}$}

\newcommand{\flux}         {ergs cm$^{-2}$ s$^{-1}$}

\shorttitle{A Sub-DLA with Unusual Abundances}
\shortauthors{Frye et al.}

\usepackage{graphicx}

\begin{document}


\title{A Sub-Damped Ly$\alpha$ Absorber with Unusual Abundances: \\ Evidence of Gas Recycling in a Low-Redshift Galaxy Group}


\author{Brenda L.~Frye}
\affiliation{Department of Astronomy/Steward Observatory, University
of Arizona, 933 N Cherry Ave., Tucson, AZ  85721, USA}

\author{David V.~Bowen}
\affiliation{Department of Astrophysical Sciences, Princeton University,
  Princeton, NJ 08544, USA}

\author{Todd M.~Tripp}
\affiliation{Department of Astronomy, University of Massachusetts, Amherst, MA  01003, USA}

\author{Edward B.~Jenkins}
 \affiliation{Department of Astrophysical Sciences, Princeton University,
  Princeton, NJ 08544}

\author{Max Pettini}
\affiliation{Institute of Astronomy, Madingley Rd., Cambridge CB3 0HA, UK}

\author{Sara L.~Ellison}
\affiliation{Department of Physics and Astronomy, University of Victoria, PO Box 1700 STN CSC, Victoria, BC, V8W 2Y2, Canada}


\begin{abstract}
  Using {\it Hubble Space Telescope/Space Telescope Imaging Spectrograph}
  G140M spectroscopy, we investigate an absorption-line
  system at $z$=0.07489 in the spectrum of the quasi-stellar object
  PG 1543+489 ($z_{\rm QSO}$=0.401).  The sightline passes within 
  $\rho = 66$ kpc of an edge-on $2L^*$ disk galaxy at a similar redshift, 
  but the galaxy belongs to a group with four other galaxies within 
  $\rho =160$~kpc.  We detect  \ion{H}{1} [log
  $N$(\ion{H}{1}/{\bf $cm^{-2}$})\:=\:19.12$\pm$0.04] as well as \ion{N}{1},
  \ion{Mg}{2}, \ion{Si}{2}, and \ion{Si}{3}, from which we measure a
  gas-phase abundance of [N/H]~=~$-1.0\pm 0.1$.  Photoionization
  models indicate that the nitrogen-to-silicon relative abundance is solar,
  yet magnesium is underabundant by a factor of $\approx$ 2.  We also
  report spatially resolved emission-line spectroscopy of the nearby
  galaxy, and we extract its rotation curve. The galaxy's metallicity is
  $\approx 8 \times$ higher than [N/H] in the absorber, and interestingly,
  the absorber velocities suggest that the gas at $\rho =$ 66 kpc is
  corotating with the galaxy's stellar disk, possibly with an inflow
  component.  These characteristics could indicate that this sub-damped
  Ly$\alpha$ absorber system
  arises
  in a ``cold-accretion'' flow. However, the absorber abundance patterns
  are peculiar. We hypothesize that the gas was ejected from its galaxy of
  origin (or perhaps is a result of tidal debris from interactions between
  the group galaxies) with a solar nitrogen abundance, but that subsequently mixed with
  (and was diluted by) gas in the circumgalactic medium (CGM) or group.  
  If the gas is bound to the nearby galaxy, this system may be an example
  of the gas ``recycling'' predicted by theoretical galaxy simulations. Our
  hypothesis is testable with future observations.
\end{abstract}






\section{\sc Introduction}
\label{sect_intro}

{\
Complex physics drives and regulates star formation in the inner regions 
of galaxies.  In addition, it is now recognized that a variety of processes 
on larger scales also have important functions in galaxy formation and 
evolution.  Galaxies must continue to accrete gas from the intergalactic 
medium (IGM) in order to explain their sustained star-formation histories.  
Accreting gas could be shock-heated to the virial temperature as it drops 
into a galaxy, but it has also been shown that the gas can radiatively cool 
and maintain a much lower temperature as it falls in, the so-called ``cold mode'' 
accretion \citep{Birnboim:03,Keres:05}.  The detailed physics that occurs in 
accreting gas can profoundly affect fundamental galactic properties
\citep[e.g.,][]{Maller:04,McCourt:12,Voit:17}. Conversely, outflows and
``feedback'' propelled by star formation or active galactic nuclei (AGNs; see,
e.g., Veilleux et al.~2005) can regulate galaxy growth
\citep{Dave:11a}, and there is some evidence that metal-rich outflows can
travel large distances from at least some galaxies
\citep[e.g.,][]{Tripp:11}.

Thus, it is important to understand the impact of large-scale inflows and
outflows to obtain a complete picture of how galaxies form and evolve.
However, observationally tracking the galactic inflows and outflows 
comprising important portions of the ``baryon cycle'' can be observationally
challenging -- gases flowing through the circumgalactic medium (CGM) will
often have low densities, and their emission features are expected to be
faint and difficult to detect.  Consequently, absorption spectroscopy,
which is much more sensitive to low-density gas, has become the workhorse
method for observing the baryon cycle and the CGM.  Intervening damped and
sub-damped Ly$\alpha$ absorbers (DLAs) with $N$(\textsc{H~i})
$\gtrsim 10^{19}$ cm$^{-2}$, detected in the spectra of background 
quasi-stellar objects (QSOs),
can be particularly interesting because they exhibit absorption in a wide
array of elements and ionization stages \citep{Peroux:03,Prochaska:03,Wolfe:05}
and thereby enable detailed
studies of abundances and physical conditions. From $z = 0$ to $z \gtrsim$
5, gas-phase abundances in DLAs and sub-DLAs decrease with increasing
redshift, but the scatter is large in all redshift bins ($\approx$1 dex),
presumably reflecting the different histories of the absorbing gas
\citep{Kulkarni:07, Meiring:09, Rafelski:12}.

To maximize the use of QSO absorbers for the study of galaxy evolution, it is
necessary to understand the relationships between the absorption systems
and the galaxies/environment.  In this regard, DLAs have been curiously
difficult to pin down, and they can arise in difficult-to-study objects
such as low surface-brightness galaxies and galaxies with very low
star-formation rates \citep[e.g.,][]{Bowen:01,Tripp:05,Battisti:12,
  Fumagalli:15}. Low-redshift absorbers are advantageous for probing
absorber-galaxy relationships because their environments can be
investigated with exceptionally deep imaging and spectroscopy, and methods
such as 21 cm emission mapping can be applied
\citep[e.g.,][]{Rosenberg:06,Borthakur:11,Borthakur:14,Borthakur:15,
  Chengalur:15,Burchett:16b,Peroux:17,Kanekar:18}.  Such studies have
  demonstrated that absorbers in galaxy group environments are relatively
  common \citep{Bielby:17, Peroux:17, Borthakur:18, Klitsch:18} but there
  is also evidence that the CGM is removed (or is highly ionized) in 
  some high-density environments \citep{Johnson:14, Burchett:16b, Burchett:18}.

One such low-redshift absorber is the low-$-z$ sub-DLA at $z \simeq 0.075$ in the
spectrum of the PG 1543+489 ($z_e = 0.40$).  This absorption system has some
puzzling characteristics and warrants follow-up observations for several
reasons. Two intervening galaxies also at $z = 0.075$ originally identified close to the QSO
sightline \citep{monk86} have relatively large impact parameters $\rho$ of 64 and
119~kpc, but surprisingly the first absorption lines detected at the
redshifts of these galaxies were the optical \ion{Ca}{2} H \& K lines
\citep{Bowen:91}.  With an ionization potential of only 11.87 eV,
\ion{Ca}{2} is easily photoionized in many contexts, and the detection of
\ion{Ca}{2} H \& K absorption usually implies a significant column density
of \ion{H}{1} in the absorption system
\citep{wild07,zych07,nestor08}. However, high $N$(\ion{H}{1}) values are
not expected at large galactocentric distances, so the detection of \ion{Ca}{2} was intriguing.
Follow-up ultraviolet (UV) observations with the Goddard High
  Resolution Spectrograph \citep[GHRS;][]{Brandt:94} on the {\it Hubble
  Space Telescope} (HST) by \citet{Bowen:95} targeted the
\ion{Mg}{2}~$\lambda\lambda 2796, 2803$ absorption doublet. These lines
were successfully detected, had moderately strong equivalent widths, and
exhibited simple component structure. At the time, Bowen~et~al.\ suggested
that the \ion{Ca}{2} and \ion{Mg}{2} absorption might arise in tidal debris
lying between the two nearby galaxies, or that a much fainter dwarf galaxy
lay hidden directly along the sightline, obscured by the glare of the
background QSO. They also considered that the absorption might be related
to the disk of the nearest, highly inclined galaxy, because the projection of
its disk along the major axis aligns closely with the position of the QSO
sightline --- even though the \ion{H}{1} in the disk was not expected to
extend out to 64~kpc at the levels of $N$(\ion{H}{1}) suggested by the
metal-line absorption.  Since these results were published, research into 
the CGM of galaxies has advanced other possible origins for absorption lines, 
such as warm outflows arising from bursts of star formation (or AGN) activity 
\citep[e.g.,][]{Strickland:04, Weiner:09, Rupke:11, Tripp:11, Chisholm:15, 
Muzahid:15}, cold flows from the IGM \citep{Keres:05, 
Dekel:09a, Ribaudo:11, Bouche:13, Rahmani:18b}, or large gas disks that may 
be corotating with the stellar disks of galaxies \citep{Steidel:02, Ho:17}.

\begin{deluxetable*}{lcccCcccc}
\tablecaption{Galaxies in the Field of \qso 
\label{tab_zs}}
\tablecolumns{9}
\tablehead{
\colhead{SDSS} & \colhead{APO} &                            &\colhead{$r$}    & \colhead{$M_r$} & \colhead{$L_{gal}$}   & \colhead {$M^*_r$}                &  \colhead{$\rho$}  &  \colhead{$\rho$} \\   
\colhead{ID}      &  \colhead{ID}   & \colhead{$z_{gal}$} &\colhead{(mag)} & \colhead{mag}    & \colhead{$(L^*)$}    & \colhead {($\log$[$M_\odot$])}  &  \colhead{(\arcsec)} &  \colhead{(kpc)}
}
\colnumbers
\startdata
\\
\hline
\hline
J154527.12+484642.2 &     G1 &   0.0751 &   16.7 &  -20.9 &   2.0 &  10.5 &    0.76 &    64 \\
J154534.68+484702.3 &    G86 &   0.0754 &   19.0 &  -18.7 &   0.3 &   9.7 &    1.15 &    98 \\
J154526.59+484453.8 &     G2 &   0.0753 &   16.0 &  -21.7 &   4.0 &  10.8 &    1.39 &   119 \\
J154533.92+484754.1 &    G96 &   0.0750 &   19.9 &  -17.7 &   0.1 &   9.3 &    1.85 &   158 \\
J154528.52+484759.4 &   G142 &   0.0759 &   17.9 &  -19.7 &   0.7 &  10.1 &    1.86 &   160 \\
J154534.48+484814.0 &    ... &    0.0      &   17.4 &    ...        &  ...  &  ...     &     2.20  &  ...    \\
J154535.86+484814.0 &    G74 &   0.0968 &   18.3 &  -19.9 &   0.8 &  10.1 &    2.28 &   245 \\
J154519.31+484808.9 &   G221 &   0.1536 &   18.1 &  -21.2 &   2.6 &  10.6 &    2.69 &   430 
\enddata
\tablecomments{Columns: (1) SDSS ID from SDSS DR14; 
(2) galaxy name adopted in this paper based on the nomenclature of the APO survey in 2002; 
(3) galaxy redshift. For the first three galaxies, G1, G86, and G2, DR14
redshifts are listed. The rest were measured at APO.
(4) Galaxy de-reddended $r$-band magnitude of the galaxy from DR14;
(5) absolute $r$-band magnitude;
(6) galaxy luminosity assuming $M_r^* = -21.2$ at $z=0.1$, from \cite{bell03};
(7) stellar mass assuming color corrections from \cite{lopez-sanjuan18};
(8) and (9) impact parameter of galaxy from QSO sightline, in arcsec and (proper) kpc.
N.B.\/ This table lists all the galaxies with known redshifts in the group at $z\simeq0.075$
within 200~kpc of the QSO sightline.
}
\end{deluxetable*}

In this study, we present new spectroscopic observations of the absorption
system toward \qso\ with HST and additional ground-based observations of
objects in the field of \qso. The HST spectra were acquired with the 
  Space Telescope Imaging Spectrograph (STIS), and were designed to
measure the metallicity and ionization structure of the absorbing gas,
while the optical data were obtained to better understand its origin. In
addition, \qso\ itself is an interesting object, being one of the most
luminous ($M_B \:= \: -26$) radio-quiet narrow line QSOs in the low-redshift
universe \citep{SG:83} with a gas outflow that extends 1150
km~s$^{-1}$ from the systemic redshift of the QSO \citep{Aoki:05}. Some of
our data may provide insight on the nature of the QSO, although in this
paper we focus on the foreground absorption system noted above.

This paper is organized as follows: in \S\ref{sect_galaxydata} we present a
synopsis of the available information on the galaxies in the field of \qso,
and present new ground-based imaging and spectroscopic data of the
galaxies. In \S\ref{sect_analysis} we show the HST STIS spectra of \qso\
and new absorption lines from the intervening absorption
system. \S\ref{sect_cloudy} discusses the abundances of the absorbing gas
derived from the new HST data and the ionization models required to better
constrain the abundances. Finally, in \S\ref{sect_discussion} we summarize
our results and discuss their implications.  Throughout the paper, we adopt
a cosmology with $H_0$ = 70 km s$^{-1}$Mpc$^{-1}$, $\Omega_M$ = 0.3,
and $\Omega_{\Lambda}$ = 0.7.

\section{Ground-based observations of the \qso\ field} 
\label{sect_galaxydata}

The field around \qso\ is covered by the Sloan Digital Sky Survey
(SDSS), which provides some relevant spectroscopic galaxy redshifts as well
as accurate broadband magnitudes. To investigate the connections between
the absorption system at $z = 0.075$ and nearby galaxies, we make use of
information from SDSS Data Release 14 (DR14) as well as our own
spectroscopic observations of galaxies in the \qso\ field.  Prior to SDSS
spectroscopic observations of this field, in 2002 June we collected
spectroscopic redshifts of galaxies near the QSO sightline using the 
  Dual Imaging Spectrograph (DIS) on the Apache Point Observatory (APO)
3.5m telescope. We used a 1.5 arcsec slit with the B300 and R300 gratings
to cover a wavelength range of $\sim 3400-7700$~\AA, and we measured the
redshifts of galaxies from their emission lines. Later, some of these
galaxies were also observed as part of the SDSS spectroscopic galaxy survey
\citep{strauss02}.  Information on galaxies from our APO program that we
use in this paper (in addition to the SDSS DR14 data) is provided in
Table~\ref{tab_zs}, including spectroscopic redshifts, SDSS $r$-band
magnitudes, angular and physical separations of the galaxies from the QSO
sightline, and estimates of the luminosities and stellar masses of the
galaxies.

We also obtained broadband images of the QSO field with SPIcam on the APO
3.5m telescope in the $R$- (50 minutes total exposure), $g$- (80 minutes)
and $i$-bands (93 minutes) in 2002 and 2003. Figure~\ref{fig_field} shows a
false-color composite image constructed by combining the SPIcam data from
the three bands.  This image is deeper than the SDSS images in similar
passbands and reveals some important morphological details that cannot be 
discerned in SDSS images (see below).  Nevertheless,  the photometric 
calibration of the SDSS data is better, so when available, we prefer SDSS 
photometry for the well-detected bright galaxies that are relevant to this study.

\begin{figure*}
\includegraphics[width=\textwidth]{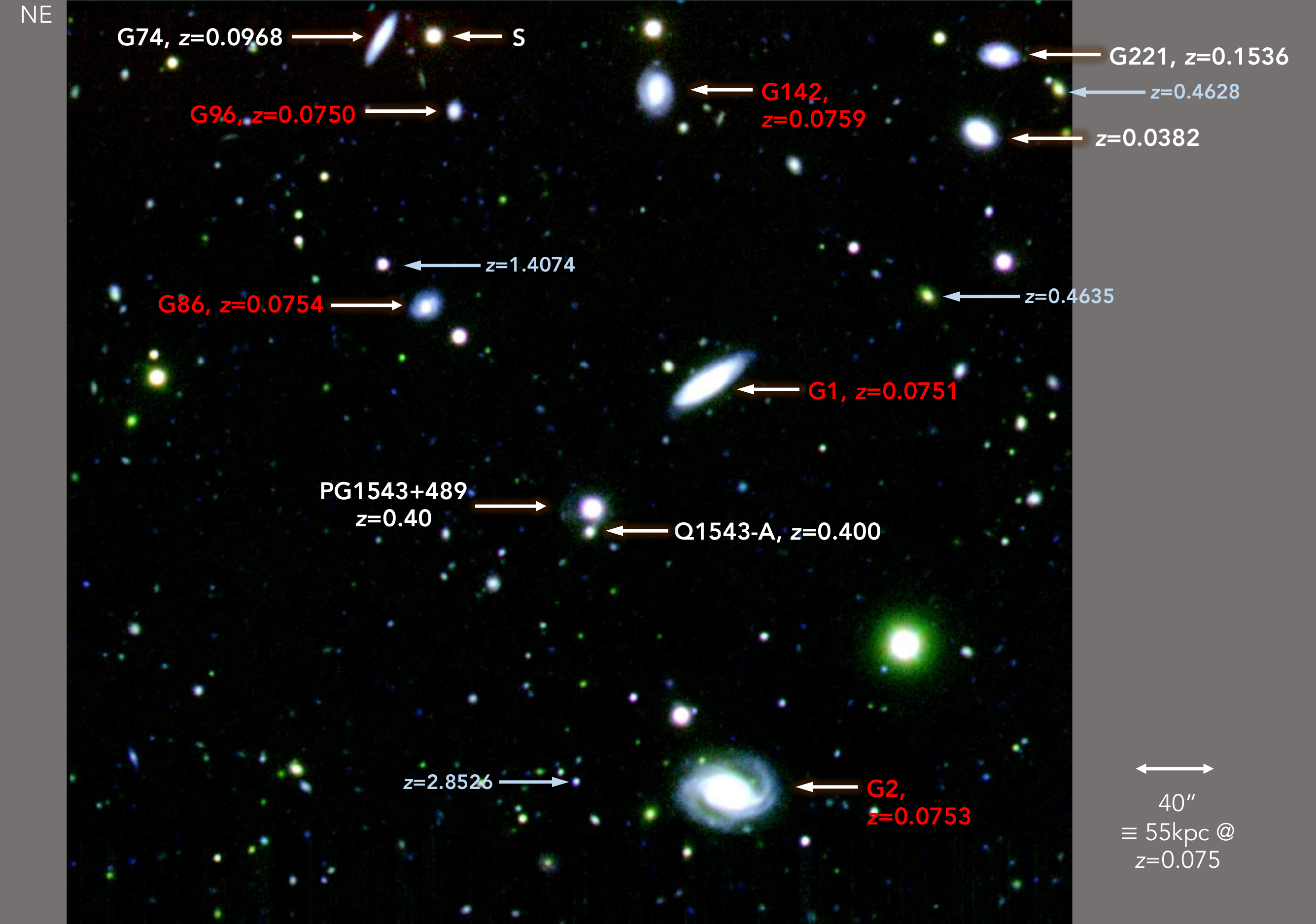}
\caption{ 
False-color image of the field of \qso\ using $g$-, $R$-, and $i$-band
images taken with the APO 3.5m telescope. All objects with known
spectroscopic redshifts, either from DR14 or our APO spectroscopic
observations, are labeled. The background QSO at $z=0.40$ is
shown, as well as the five galaxies at $z\simeq 0.075$ lying within an impact
parameter of $\simeq$200~kpc from the QSO sightline (labeled in red) and
one star (`S').  
Objects that are at or in front of the QSO but are otherwise not relevant
to this study (because the objects are at significantly different redshifts
from the absorption-line system) are annotated with white labels, 
and small blue labels mark objects that are behind the QSO at higher redshifts. 
The field of view is $7.2 \times 5.3$ arcmin.
\label{fig_field}}
\end{figure*}

\begin{figure}[t]
\begin{center}
\includegraphics[width=0.475\textwidth]{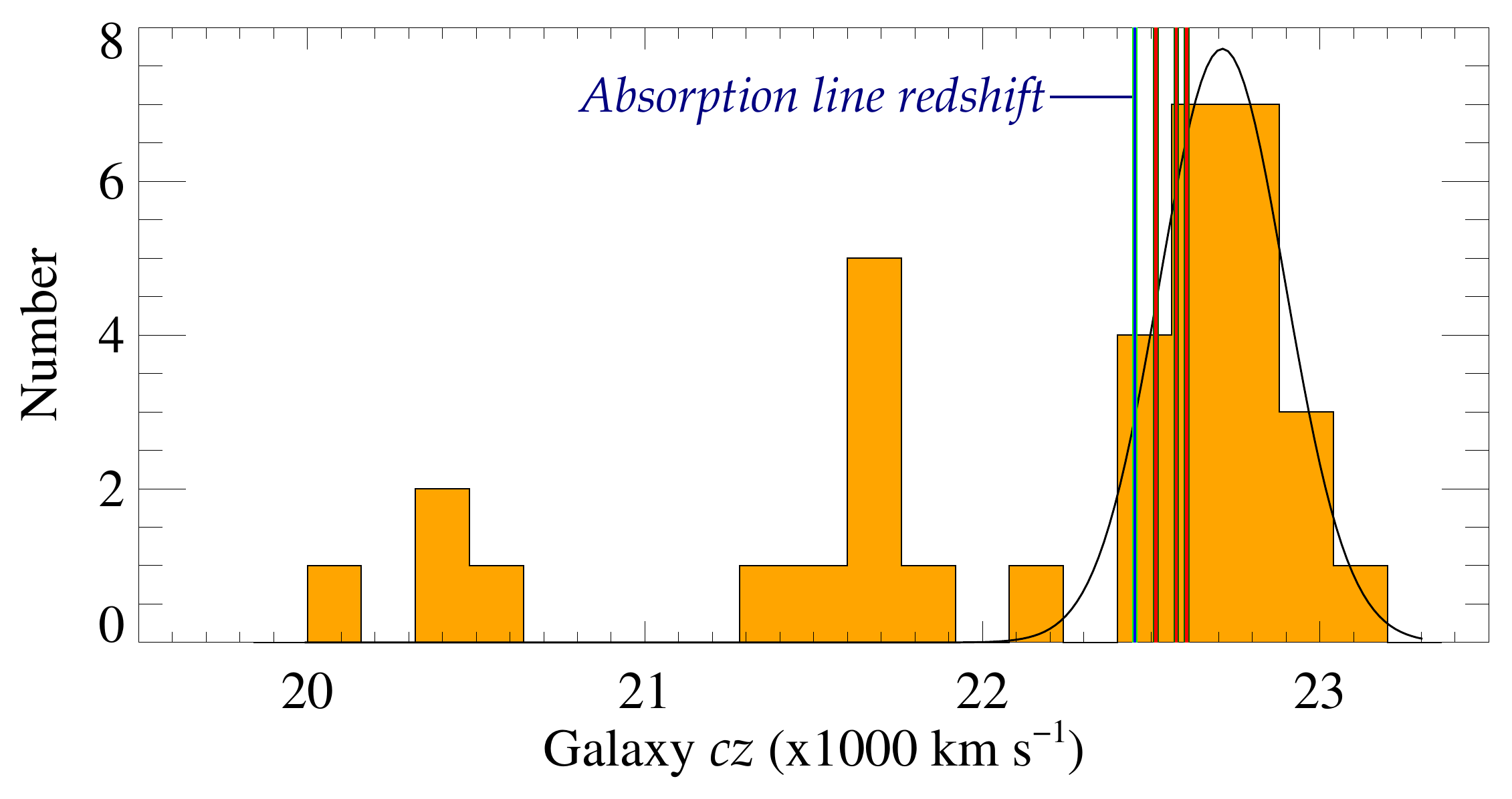}
\caption{Redshift distribution of galaxies from DR14 and our own 
redshift survey within $0.6\degr$ of the sightline to \qso , or 
$\simeq$3.1~Mpc. The absorption-line redshift at $z=0.07489$ is 
indicated by the blue vertical line, while the redshifts of the three 
central galaxies are marked with red vertical lines (from left to right, 
G1, G2, and G86).  A Gaussian fit to the profile gives the center of 
the distribution to be $22,710\,\pm\,30$~\kms, with a width of 
$\sigma \sim 190\,\pm 30$~\kms . Although not shown in this figure, 
these galaxies are spread roughly uniformly in two dimensions 
within the $0.6\degr$ radius.
  \label{fig_lss}}
\end{center}
\end{figure}

\subsection{The galaxy group affiliated with the \qso\ sub-DLA}

The brightest foreground galaxies lying near the sightline
toward \qso\ are at
impact parameters $\rho$ = 64 and 119 kpc. The SDSS designations for both
are listed in Table~\ref{tab_zs}, but in the rest of this paper, we refer to
the two as ``G1" and ``G2," in keeping with their original
designations by \cite{monk86}.  Our APO survey has revealed a third ---
albeit fainter --- galaxy with a redshift similar to those of G1 and G2 and
a comparable impact parameter of $\rho$ = 98 kpc (G86; see Table~\ref{tab_zs}).
 In addition, the APO survey located two other galaxies at
similar redshifts with $\rho$ = 158 and 160 kpc (G96 and G142 in
Table~\ref{tab_zs}).  Thus, a group of five galaxies, all within an impact
parameter of 160 kpc from the sightline, is found at the absorber redshift.
These galaxies are identified with red labels in Figure~\ref{fig_field}.
Objects with spectroscopic redshifts that are significantly different from
the group redshift are also indicated in Figure~\ref{fig_field}
with white or blue labels.

SDSS provides additional information on galaxies at angular
separations from the sightline larger than the angular limit of our APO survey. To
investigate the larger scale extent of the group at $z=0.075$, we collated
all galaxy redshifts available from SDSS DR14 within a radius of 122
arcmin, or 10~Mpc at the group's redshift. As our sample was drawn
from SDSS, we have a similar 
limiting magnitude of $r\,\simeq\,17.8$,  or $M_{r} = -19.9$ (i.e., $0.1\,L_r^*$, 
at $z=0.075$), 
although some of the galaxies observed at APO
are fainter than this limit.  The two-dimensional distribution of the galaxies
within the range $20,000 \leq cz \leq 24,000$ displays no clear large-scale
structural features (e.g.,~galaxy filaments) or any association with, e.g.,~a
denser cluster, but a group of galaxies appears to be confined to a radius
of $\approx 0.6\degr$, or 3.1~Mpc, around the QSO sightline, at the group
redshift of $\sim 0.075$. The one-dimensional distribution of these galaxies is shown in
Figure~\ref{fig_lss}. The peak of the distribution occurs at
$22,710 \pm 30$~\kms, with the three galaxies G1, G86, and G2 all offset to
slightly lower velocities (see the red vertical lines in
Figure~\ref{fig_lss}). The group consists of 21 galaxies within
$\pm 500$~\kms\ of $z=0.075$, and apart from G1 and G2, only two other
galaxies are obviously disks, with the rest being of early or S0
type. It is possible that G1 and G2 are therefore in their early stages of
falling into a rich group of galaxies.
   
\subsection{Properties of G1}

In order to obtain spatially resolved information on G1 (e.g., its rotation curve), we again
used DIS at APO, this time with a 1.5 arcsec slit aligned along the major
axis of the galaxy. The data were taken on  2005 March 11 with the B300/R300
gratings, and we obtained five 1200 s exposures at high air-mass. The
spectra were binned on-chip to give a spatial dispersion of 0.84 arcsec
pix$^{-1}$, and a spectral resolution of 300~\kms\ FWHM.  The five
individual frames were debiased, flat-fielded, and then coadded to
produce a final 2D frame. Each row (spectrum) was wavelength-calibrated
using the same row from an arc lamp exposure taken prior to the start of
the observations. Finally, the data were flux-calibrated using the standard
star BD+33d2642.

\subsubsection{Metallicity and star-formation rate of G1 \label{sect_gal_z}}

Emission lines of [\ion{O}{2}]~$\lambda 3727$, 
[\ion{O}{3}]~$\lambda \lambda$4959, 5007, H$\beta$, H$\alpha$, and 
[\ion{N}{2}]~$\lambda 6583$ were detected from many of the individual CCD
rows comprising the galaxy spectrum.  The total H$\alpha$ flux 
from G1
summed over all these rows is $\simeq 1.0 (\pm0.2) \times
10^{-14}$~\flux. This includes all the flux from the disk of the galaxy,
but does not account for losses from the galaxy being slightly wider than
the slit or from internal extinction, which we cannot measure due to the
non-detection of other Balmer lines.  H$\alpha$ decreases as galactocentric
distance $r$ increases and is depressed in the inner $2.5-4.2$ arcsec,
unlike [\ion{N}{2}] which shows no such dip at small $r$.

The H$\alpha$ and the [\ion{N}{2}] lines enable us to measure the gas-phase
metallicity of G1 using the strong-line N2 index, defined as N2 =
$\log$([\ion{N}{2}]~$\lambda 6583$\:/\:H$\alpha$).  N2 is largely constant
across the disk except in the inner region where H$\alpha$ decreases. This
may indicate a change in ionization by a non star-forming
source, such as a low-luminosity AGN at the galaxy's center. If we
ignore this region, then the lower limit to the H$\alpha$ flux 
is $7.5\,(\pm 0.3) \times 10^{-15}$~\flux. The
resulting total H$\alpha$ luminosity is $\log $[H$\alpha$~(ergs
s$^{-1}$)]$\:= 41.02\pm0.01$, which gives a star formation rate of
$0.8\pm0.03$~$M_\odot$~yr$^{-1}$ using the relationship given by
\citet{kennicutt98a}.

The average N2 ratio for the two outer regions is similar,
$-0.32\pm0.04$, which gives a metallicity of $\simeq 8.60\pm0.02$ using the
calibration of \citet{marino13_ra},\footnote{As is customary for interpreting metallicity measurements based on the analysis of
emission lines, we note that electron-temperature-based empirical
calibrations produce metallicities that are $\sim 0.2-0.4$ dex lower than
strong-line methods based on photoionization models, such as the N2 index
used herein \citep[e.g.][and references\/ therein]{kewley08,LopezSanchez12}.} or [O/H] = $-0.10\pm0.05$, i.e.
$\sim 0.8$ times the solar value assuming 12\,+\,log[O/H]$_\odot$ =
$8.69\pm0.05$ \citep{Asplund:09}.  We note that this is only 0.1 dex smaller than that
derived using the original N2 calibration of \citet{pettini:04}, whose
sample was more sparse than the data available to \citet{marino13_ra}. The
measurement errors on N2
for individual spatial pixels are larger, $\approx$$\pm 0.1$~dex, and there is no
compelling evidence for a change in N2 as a function of galactocentric
radius --- as might be expected from a metallicity gradient in the disk ---
given the large errors.  We return to this measurement in \S
\ref{sect_discussion}.

\subsubsection{Rotation curve of G1}

\begin{figure}
\begin{center}
\includegraphics[width=0.5\textwidth]{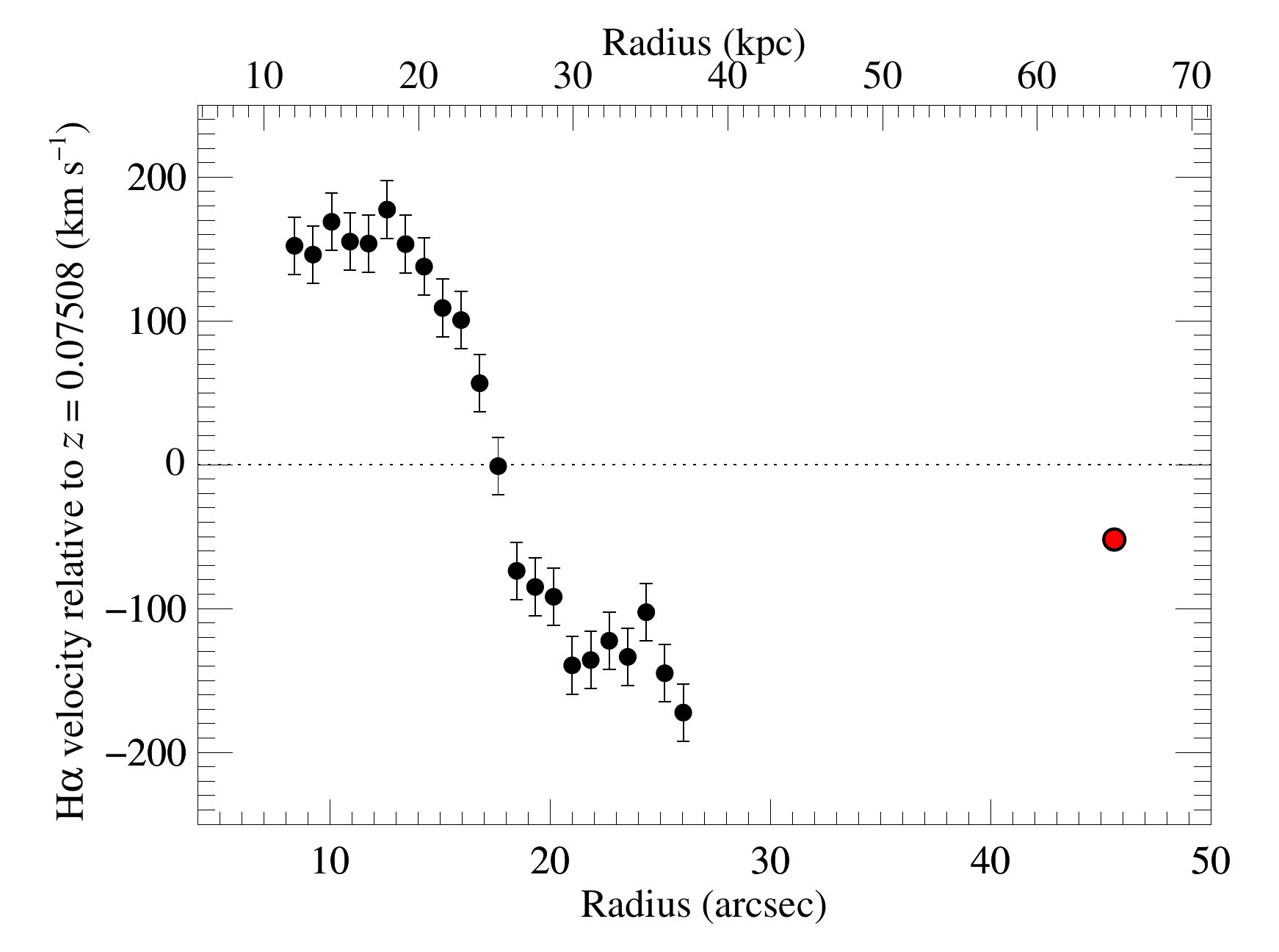}
\caption{The rotation curve for galaxy G1 toward \qso, determined from
  H$\alpha$ emission lines (black points). The velocity of the absorption
  system at $z=0.07489$ fit to the H I Ly$\alpha$ line is shown as a red circle.
\label{fig_newrot}}
\end{center}
\end{figure}

Having extracted 1D spectra from individual CCD rows of our APO data, and
with the slit oriented along the major axis of G1, we were able to
construct a rotation curve of the galaxy from the emission lines (Figure~\ref{fig_newrot}). 
The seeing at the time was $1-2$ arcsec FWHM, so each
row of the CCD (0.83 arcsec pix$^{-1}$) corresponds to a sampling of approximately half a
resolution element. 
We
measured the velocity in two ways, first by  fitting Gaussian
profiles to the H$\alpha$ and [\ion{N}{2}] lines, and then by using the
\textsc{fxcor} routine in IRAF, which computes radial velocities between
spectra via Fourier cross-correlation. With the latter method, we used the
1D spectrum at the center of the galaxy as the reference spectrum.  Both
methods gave similar results, to within the errors generated by the
\textsc{fxcor} routine.  We return to the relationship between the velocity of
the disk and the absorption detected toward \qso\ in \S\ref{sect_discussion}.

\subsection{A search for faint galaxies close to the sightline of \qso}

With our prior knowledge of the peculiar nature of the absorption system
toward \qso , we were motivated to search for other faint galaxies 
close to the sightline that could be the source of the
absorption rather than the three galaxies close to the sightline listed in
Table~\ref{tab_zs}. A closer and more detailed representation of the APO
composite image of the QSO field is shown in Figure~\ref{fig_dvbimage}.
Apart from the bright satellite galaxy to the
south of the QSO (labeled Q1543$-$A), which is known to be
at the redshift of \qso\ \citep{Bowen:95}, there is additional nebulosity
directly to the east of the QSO.  In this section, we discuss data which we
obtained in order to measure the redshift of this `fuzz.'  In principle, this fuzz could be the
hypothesized galaxy close to the sightline that is the source of the
absorption-line
system, but the host galaxies of QSOs are often readily apparent in images
of this depth, so this could, alternatively, simply be light from the QSO
host galaxy.  As we show below, all available evidence indicates that the
fuzz is indeed the host galaxy of the QSO and is unrelated to the
absorption system.

\begin{figure}
\includegraphics[width=0.5\textwidth]{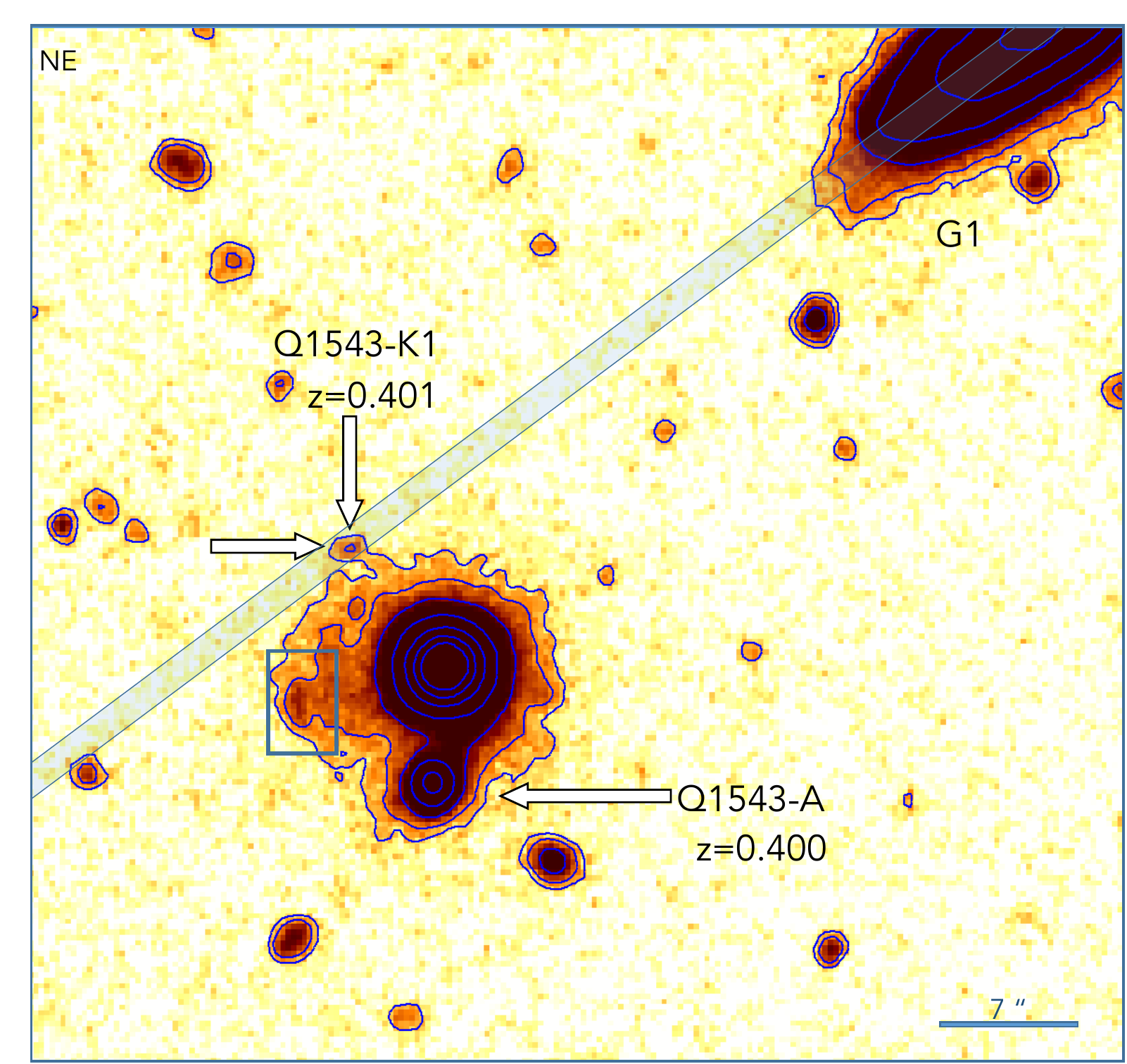}
\caption{Image of the QSO PG1543+489, which has a redshift of 
$z=0.401$.  This figure shows the $g$-, $R$- and $i$-band images added
 together to improve the signal-to-noise ratio of low surface brightness features. 
The sightline to the QSO passes through the CGM of galaxy G1.
A small satellite, labeled  Q1543-A, has the same redshift as \qso\ \citep{Bowen:95} 
and lies 6$\farcs$0 or 32~kpc, away on the plane of the sky.  
Interestingly, a 
combination of clumpy and diffuse emission is seen around the QSO out to large 
radii of $\sim 8^{\prime \prime}$ toward the east. We have measured a redshift 
of $z=0.40$ for one particular clump, at a separation of 
7$\farcs$3 or 40~kpc, labeled here as Q1543--K1.  
The DIS slit used to measure the rotation curve of G1 and detect the
emission from K1 is shown in blue.
The blue box shows the field of view of our GMOS-N IFU observations. 
The orientation is the same as in Figure~\ref{fig_field}.
\label{fig_dvbimage}}
\end{figure}

Fortuitously, the orientation of the DIS slit used to measure the rotation
curve of G1 also passed through a bright knot in the extended fuzz surrounding
the QSO, which we label as Q1543--K1
in Figure~\ref{fig_dvbimage}.} Q1543--K1 has a $g$-band magnitude of
$24.2\pm 0.9$ and lies 7.3 arcsec from the QSO sightline. At the position
of Q1543--K1, our APO spectrum shows clear emission lines of H$\alpha$,
[\ion{O}{3}] $\lambda \lambda$4959,5007, H$\beta$, and
[\ion{O}{2}] $\lambda$3727 at a redshift of 0.401, i.e., the same as the
QSO redshift and satellite Q1543--A.

In addition to the APO data, we used the Gemini Multi-Object Spectrograph
North (GMOS-N) IFU unit in the optical to search for emission lines at the
redshift of the QSO or that of the absorption-line system. We acquired nine
1200 s integrations in June and July 2016 centered on the most easternmost
part of the fuzz, with a $3.5\times 5$ arcsec field of view shown as a blue
box in Figure~\ref{fig_dvbimage}. Our observations used the grating to
give a central wavelength of 7000 \AA \ and a coverage of $5618 - 6922$
\AA\ at a measured dispersion of 0.47 \AA\ pix$^{-1}$.  In this observing
mode, the data read out into an array of 33 $\times$ 49 spatial pixels
(spaxels) for each wavelength interval.  We reduced the data using the
IRAF/Gemini software pipeline, with the addition of the \textsc{lacosmic} task which
ensures a robust removal of cosmic rays \citep{vanDokkum:01}.  The
IRAF/Gemini set of tasks does not include a means to coadd multiple 3D data
cubes (in our case, nine separate integrations), so we created a separate
module for that task.

The final coadded data cube showed no evidence for any emission lines from
the QSO or from the intervening absorption-line system.  Binning the data
spatially by up to a factor of 8 failed to show any signal.  We note that 
the only grating option available at the time of the observations was R400, which 
covers only the H$\beta$ emission line at the redshift of the QSO, and no emission 
lines at the redshift of G1.  The non-detection of H$\beta$ is unfortunate, and also 
not so surprising, as the H$\beta$ emission line is already established to be weak in
 our spectrum of the Q1543--K1 fuzz component.

The presence of a well-detected companion (Q1543--A) to the south of the QSO
suggests that these two objects could be interacting.  Such interactions
often produce extended nebulosity similar to the `fuzz' in
Figure~~\ref{fig_dvbimage}, and because the brightest spot in the fuzz has
the same redshift as the QSO, it is most likely that the fuzz is due to
material in the immediate vicinity of \qso\ and is not a foreground galaxy
affiliated with the absorption-line system.

\section{High-resolution ultraviolet spectroscopy of
  \qso}
\label{sect_analysis} 

To expand on our set of the \ion{Ca}{2} H \& K and \ion{Mg}{2}~$\lambda\lambda 2796, 2803$ 
ions for analysis of the absorption-line system, we reobserved \qso\
with STIS on 13 October 2001 using the G140M grating and the 52$^{\prime \prime}\,\times\,$0$\farcs$1
aperture.  This STIS grating has a spectral resolving power
$R$\,$\approx$\,10000, i.e., spectral resolution $\approx$ 30 km s$^{-1}$
\citep{Woodgate:98}.  Two wavelength settings were used, with central
wavelengths of 1272 and 1321~\AA , for a total exposure of 189 minutes.  
After applying the standard CALSTIS pipeline to extract the one-dimensional
spectra, we coadded the overlapping regions of the two
spectra with a weighting based on the inverse of each spectrum's variance.
This produced a single spectrum covering the $1245-1349$ \AA\ wavelength
range.

Several resonance transitions of important species at the redshift of the
absorption-line system are covered in this observed spectrum, including the \ion{H}{1}
Ly$\alpha$ line; the \ion{N}{1}~$\lambda \lambda$1199.55, 1200.22 and
1200.71 triplet; \ion{Si}{2}~$\lambda \lambda$1190.42, 1193.29;
\ion{Si}{3}~$\lambda$1206.50, \ion{N}{5}~$\lambda \lambda$1238.82, 1242.80;
and two of the three lines of the \ion{S}{2} triplet at 1250.58 and 1253.81
\AA .  Of these target lines, the new STIS spectrum clearly revealed the
\ion{H}{1} Ly$\alpha$, \ion{N}{1}, \ion{Si}{2}, and \ion{Si}{3} lines at
high significance.  The \ion{S}{2} and \ion{N}{5} features are not apparent
above the noise in the new data.  We note that low-resolution spectra
of \qso\ have also been recorded with the Faint Object Spectrograph and the
Cosmic Origins Spectrograph \citep[e.g.,][]{Borthakur:13}. Unfortunately,
due to the low spectral resolution, these additional data are much less
sensitive and generally do not place sufficiently stringent limits on metal
absorption lines to be useful for the purposes of this paper.

We continuum-normalized the STIS data using the methods outlined by
\citet{Sembach:92} (see below), and show the normalized absorption profiles of
the detected lines in Figures 5 and 6.  We also reanalyzed
the \ion{Mg}{2} lines observed with GHRS by \citet{Bowen:95}, using the
archival data but reprocessed with the final version of the CALHRS
pipeline, v1.3.14 (2004), which includes several iterations of pipeline
improvements implemented after the calibration of the data shown in the
original paper.  The re-reduced and continuum-normalized \ion{Mg}{2}
profiles are also shown in Figure~\ref{fig_spec}.

\begin{figure*}
\includegraphics[width=\textwidth]{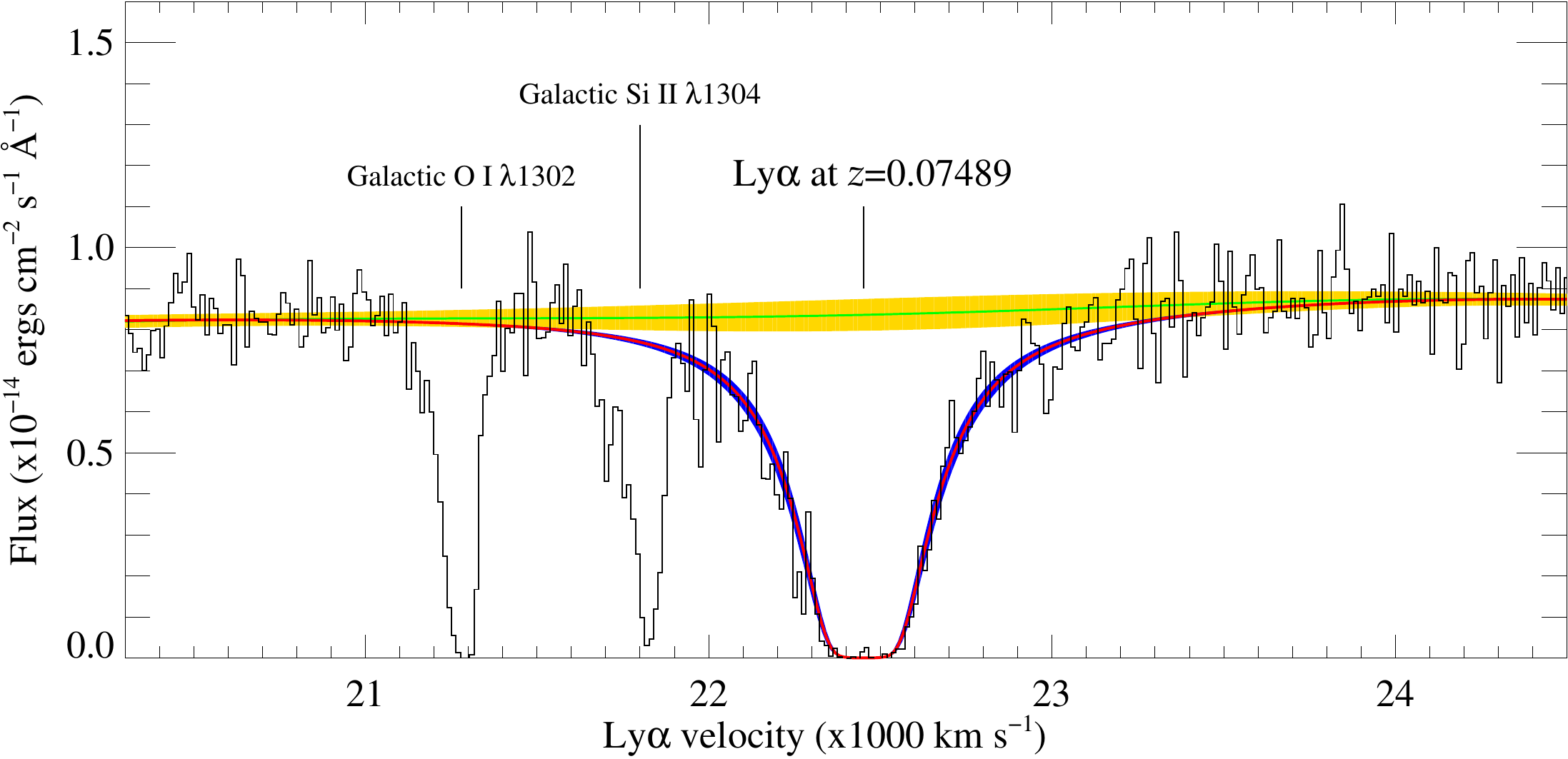}
\caption{Portion of the STIS spectrum of PG~1543+489 showing strong Ly$\alpha$
  absorption at $z=0.07489$. The fit to the continuum is shown as a green
  line, while the range bounded by the upper and lower continuum fits used to derive the error
  in $N$(\ion{H}{1}) is shown in yellow. The best Voigt profile fit to the Ly$\alpha$
  line is drawn in red, while the region bounded by $\pm1\sigma$
  differences in  $N$(\ion{H}{1}) is shown in blue.  
  }
\end{figure*}

\begin{figure}
  \centering
  \includegraphics[width=0.5\textwidth]{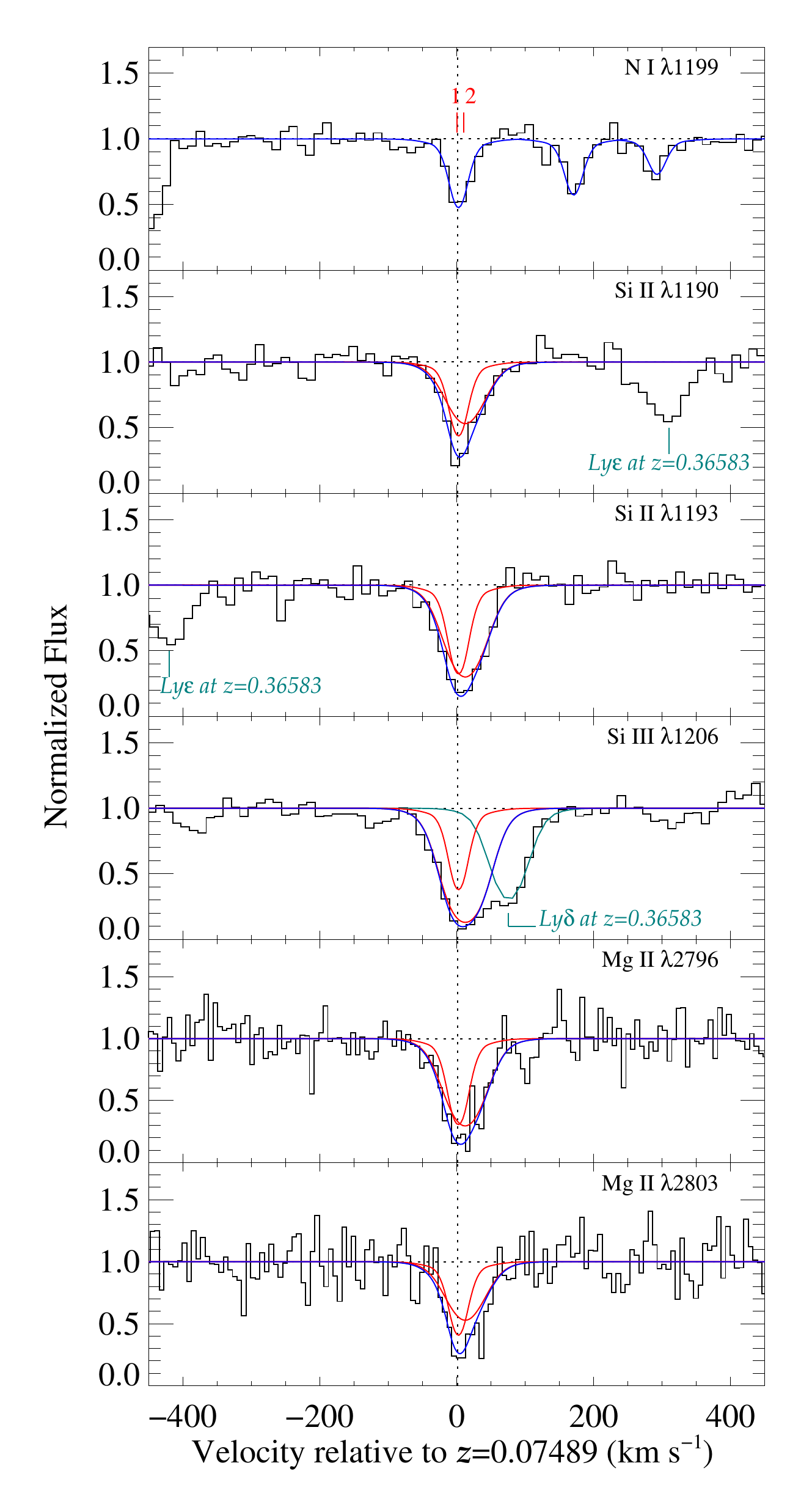}
  \caption{Velocity plot of the STIS and GHRS (the latter covering the
    Mg~II lines) spectra of \qso , showing the lines detected at
    $z=0.07489$. For \ion{N}{1}, the spectrum is centered on the $\lambda
    1199.5$ line, but the fits to the $\lambda 1200.2$ and $\lambda$1200.7 lines
    are also shown at higher velocity. The \ion{Si}{3}~$\lambda 1206$ line is blended with an
    Ly$\delta$ line at $z=0.3658$, but that line can be modeled (shown in
    teal) by fitting 
    Ly$\gamma$, Ly$\epsilon$, Ly$\zeta$, and Ly$\eta$ lines ($\log
    N\,$(\ion{H}{1})$= 15.37\,\pm\,0.02$, $b$\,=\,27.8$\,\pm\,1.2$~\kms) detected elsewhere in the
    data. The two components of the best-fit model are shown as red lines,
    and their resulting composite as a blue line.
    \label{fig_spec}}
\end{figure}

\begin{deluxetable}{rll}
\tablecaption{Column densities of ions and element ratios at $z$\,=\,0.07489 toward PG1543+489}
\tablecolumns{3}
\tablehead{
\colhead{ }
& \colhead{Component  1} & \colhead{Component 2} \\
\colhead{}
& \colhead{$v=2.2\pm1.0$ km s$^{-1}$} & \colhead {$v=12.2\pm0.9$ km s$^{-1}$} \\
\colhead{}
& \colhead{$b=10.2\pm1.8$ km s$^{-1}$} & \colhead {$b=31.6\pm1.6$ km s$^{-1}$} 
}
\startdata
$\log\,N$(H\,{\sc I})                       & 19.08 ($\pm$\,0.05)$^a$       & -- \\
\hline
$\log\,N$(N\,{\sc I})                       & 13.91 ($-0.05$,\,+0.07)        &  $\lesssim$ 13.0 \\
$\log\,N$(N\,{\sc V})                      & $<$\,13.5                        &  $<$\,13.5 \\
$\log\,N$(N\,{\sc I}/H\,{\sc I})         & $-5.17$ ($-$0.07,\,+0.09) &  -- \\
$Z$(N/H)$^b$                               & $-1.0$ ($\pm\,0.1$)    & -- \\
\hline
$\log\,N$(Si\,{\sc II})                      & 13.72 ($-0.08$,\,+0.22)     & 13.69 ($-0.07$,\,+0.03) \\
$\log\,N$(Si\,{\sc III})                     & 13.10 ($\pm$\,0.29)        &  13.53 ($-0.04$,\,+0.02) \\
$\log\,N$(Si\,{\sc II}+Si\,{\sc III})    & 13.81 ($-0.09$,\,+0.19)      & -- \\
$Z$(Si/H)$^b$                              & $-0.78$ ($-0.11$,\,+0.20)  & -- \\
\hline
$\log\,N$(Mg\,{\sc II})                                 & 13.40 ($-0.19$,\,+0.36)  & 13.31 ($-0.09$,\,+0.06)\\
$\log\,$(Mg\,{\sc II}/H\,{\sc I})      & $-5.68$ ($-0.20$,\,+0.36)  & -- \\
$Z$(Mg/H)$^b$                          & $-1.25$ (-0.20,\,+0.37) & -- \\
\hline
$\log\,N$(S\,{\sc II})                                     & $<\,14.5$  & $<\,14.5$ \\
\enddata
\tablenotetext{a}{The units of all column density values are in $cm^{-2}$.}
\tablenotetext{b}{These are the abundances $Z$\,=\,log[X/H]\,--\,log[X/H]$_{\odot}$
  that N, Si, and Mg would have if there were no ionization
corrections and no dust depletion in the absorber for component 1. Solar values are $\log\,N$:  $7.83 \pm 0.05$; Si:  $7.51 \pm 0.03$,
and Mg:  $7.57 \pm 0.04$, from \citet{Asplund:09}.}
\end{deluxetable}

As expected, the \ion{H}{1} Ly$\alpha$ line shows strong damping wings (see
Figure 5), and these wings provide
excellent constraints on the total \ion{H}{1} column density of the
absorber.  By the fitting the \ion{H}{1} Ly$\alpha$ line with a single
damped profile, we obtain $z_{\rm abs} = 0.07489$, which we adopt for the
systemic redshift of the absorption system.  We also obtain a total column
density (in cm$^{-2}$) of log $N$(\ion{H}{1}) = 19.08$\pm$0.05.  
This result shows that the absorption-line system falls into the category
of the sub-DLAs introduced in \S\ref{sect_intro}. 
However,
as we discuss below, the absorption profiles of the metals show clear
indications of two absorption components, which introduces a complication
in our analysis of the absorber ionization and metallicity; although the
damped Ly$\alpha$ profile provides a precise measurement of
$N$(\ion{H}{1}), this is the total column density summed over both
components.  Consequently, we will need to consider how the \ion{H}{1}
might be distributed between the two components in our ionization modeling.

To derive the velocity centroids ($v$), column densities ($N$), and Doppler
parameters ($b$) of the metal absorption lines shown in
Figure~\ref{fig_spec}, we fitted Voigt profile models, convolved with the
STIS line-spread function for the G140M grating, to the normalized data.
We fitted all eight metal lines simultaneously, allowing the individual metal
column densities $N$ to vary but requiring that all metals in a given
component have the same $v$ and $b$ values\footnote{The constraint of a
  fixed $b$-value for all of the metals implicitly assumes that the line
  widths are dominated by microturbulence within the absorber rather than
  by thermal broadening.}.  We found that while a single-component fit was
adequate for the \ion{N}{1} triplet, it could not reproduce the structure
seen in the \ion{Si}{2}, \ion{Si}{3}, or \ion{Mg}{2} lines, which clearly
exhibit an additional weaker (but blended) component redward of the
stronger one.  Our best-fit
Voigt-profile models are overlaid on the data in Figure~\ref{fig_spec}, and
the profile parameters are summarized in Table~2.  In
Figure~\ref{fig_spec}, we show the full fit (i.e., including both
components) to each profile with a solid blue line, but we also show the
models for the two discrete components with red solid lines.  In one
instance (the \ion{Si}{3} line), it was necessary also to model the
contribution from an \ion{H}{1} Ly$\delta$ line at a different redshift
($z_{\rm abs}$ = 0.36583) that is moderately blended with the \ion{Si}{3}
absorption.  Fortunately, the shape of this blended interloper is tightly
constrained because \ion{H}{1} Ly$\epsilon$, Ly$\zeta$, and
Ly$\eta$ at the same redshift are covered by the STIS data.  Consequently, this blend has little impact
on our \ion{Si}{3} measurements.  For the metal lines, uncertainties in the
profile parameters were estimated using the method of \citet{Bowen:95}.
The error that we report for the \ion{H}{1} column was derived by
considering the upper and lower bounds on the continuum from the formalism
of \citet{Sembach:92}, as implemented by \citet{bowen08}.

\begin{figure}
\hspace*{-0.7cm}\includegraphics[width=0.55\textwidth]{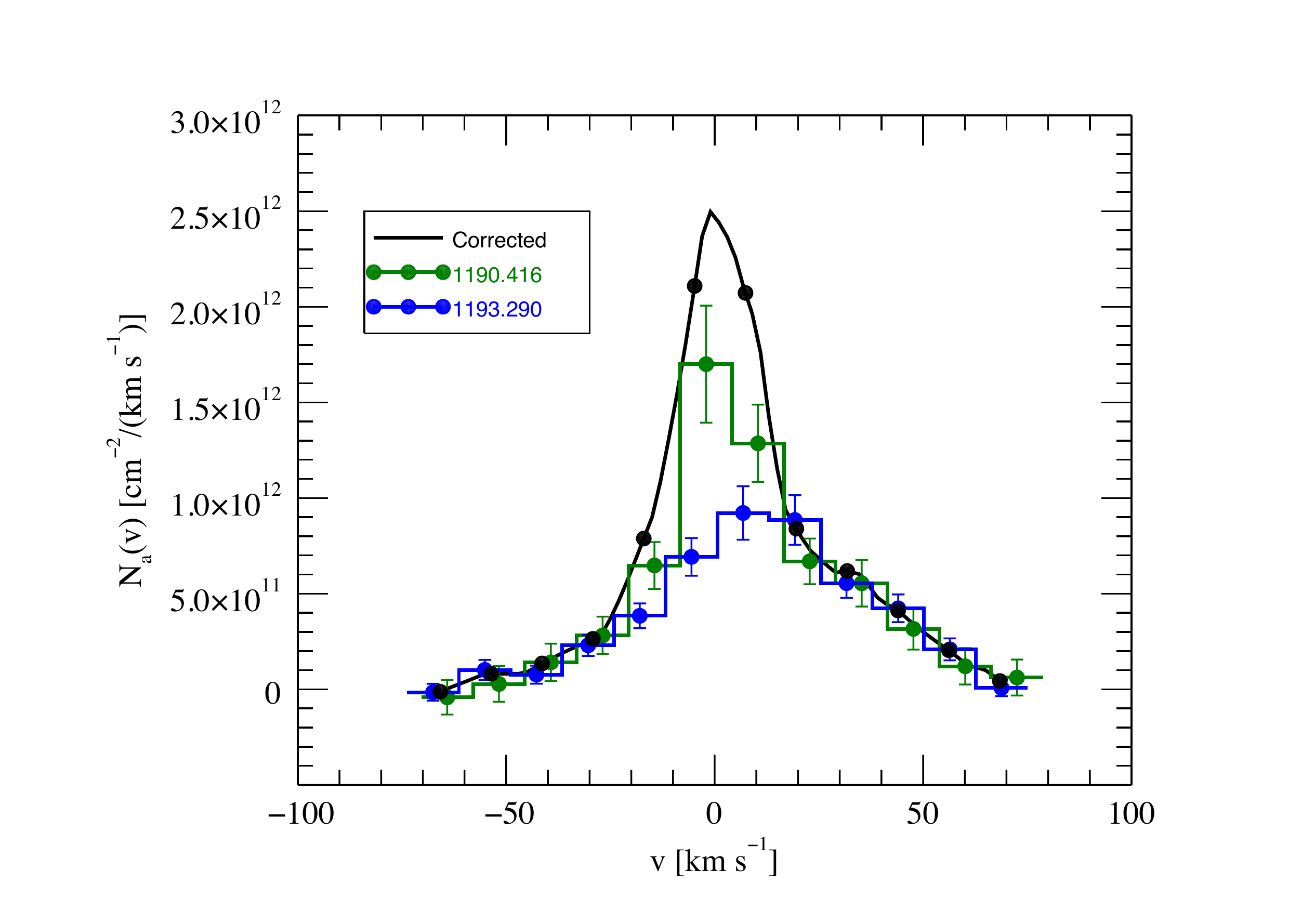}
\caption{The apparent column densities of the \ion{Si}{2} $\lambda \lambda$1190,1193 lines in the sub-damped Ly$\alpha$ absorber at
  $z_{\rm abs}$ = 0.07489 in the spectrum of \qso.  The peaks of each line
  do not match in height, indicating the presence of some unresolved
  saturation.  The corrected apparent column density, derived using the
  technique of \citet{Jenkins:96}, indicates a two-component structure that
  is consistent with our results from Voigt-profile fitting.  Integration
  of the corrected apparent column density indicates a total column density
  of $\log N$(\ion{Si}{2})$=13.96\pm 0.02$, which is in good agreement with
  our Voigt-profile line-fitting values of $14.00$($-0.07,+0.12$).
\label{AOD}}
\end{figure}

We corroborate the two-component structure that we inferred from
Voigt-profile fitting by performing an independent apparent column density
analysis \citep{Savage:91} of the \ion{Si}{2}~$\lambda 1190,1993$ lines.
This approach can also be used to reveal unresolved saturation in
components if multiple transitions with different oscillator strengths are
available.  The apparent column density profiles for the \ion{Si}{2} lines
are shown in Figure~\ref{AOD}, where it is clear that the values for
the apparent optical depth 
$N_{a}(v)$ of the two lines do not match, indicating the presence of
unresolved saturation in one or both of the lines.  As the saturation is
expected to be only moderate, it can be corrected using the prescription
given by \citet{Jenkins:96}. The result of such a correction is shown as a
black line in Figure~\ref{AOD}, and we see that two components are
indicated by direct analysis of the apparent column density profiles.  We
also find that integration of the saturation-corrected apparent column
density profile yields a total column density in good agreement with our
profile fitting results.  The same saturation is also seen in the profile
fits to the \ion{Mg}{2} lines, a result we return to in the next section.

For convenience, we hereafter refer to the two components as ``component
1'' and ``component 2'' as labeled in Table~2.


\section{Ionization Models and Absorber Abundances}
\label{sect_cloudy}

We now examine the implications of the new STIS measurements regarding the
abundances of the sub-DLA of \qso, which can provide insight into the origin
and history of the absorbing gas.  As noted above, measurement of chemical
abundances in this system is somewhat complicated by our conclusion that
there are two components in the metal absorption profiles (see previous
section), while we only have constraints on the total \ion{H}{1}
column.  To overcome this problem, we have conducted two analyses of the
absorber abundances.

\begin{figure*}
\includegraphics[width=0.98\linewidth]{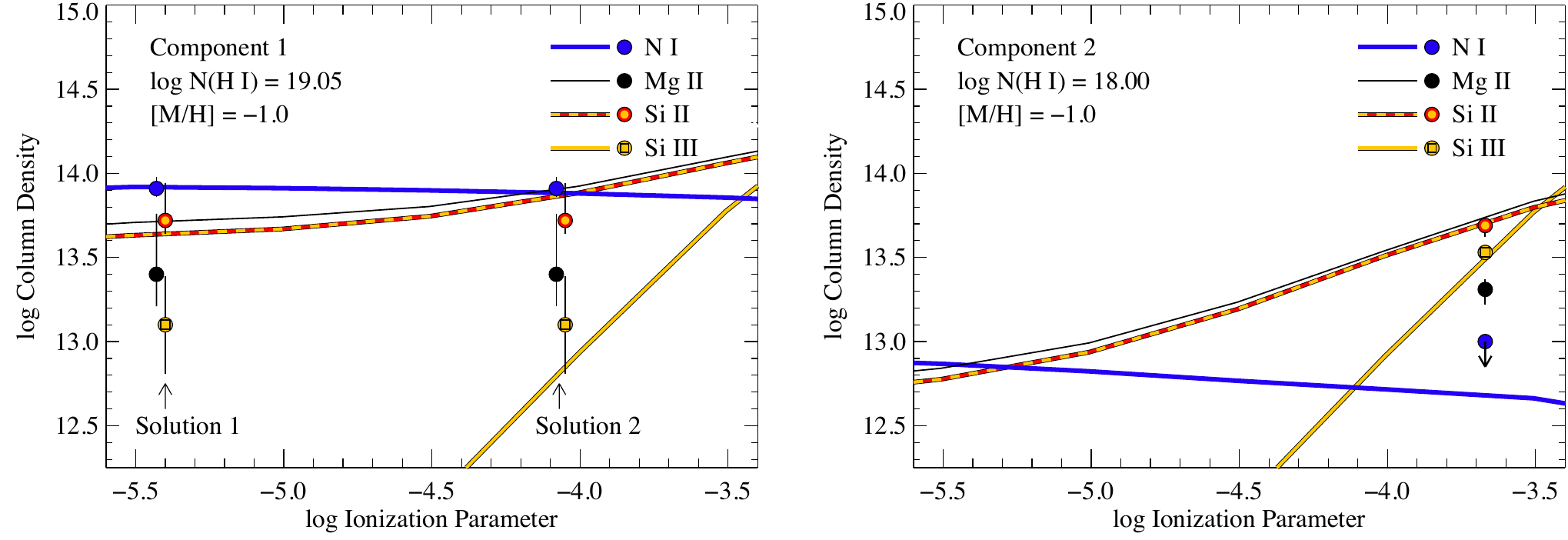}
\caption{Photoionization models fitted to the measured column densities in
  component 1 (left panel) and component 2 (right panel) of the
  sub-DLA in the spectrum of \qso.  For the models shown in this figure, we
  assume that the relative abundances are the same as those observed in the
  photosphere of the Sun \citep[from][]{Asplund:09} and there is no
  depletion of N, Mg, or Si by dust. The column densities predicted by the {\sc
    cloudy} models are shown as a function of the ionization parameter.  
Model column densities are shown with smooth curves,
  and the observed column densities (from Table~2) are indicated with
  discrete symbols. For the stronger and more neutral component 1, we discuss
  multiple ways that the observed column densities could be interpreted in
  the context of this model; the two options that we discuss are labeled as
  ``Solution 1'' and ``Solution 2'' in the left
  panel. \label{fig_cloudy_nodepl}}
\end{figure*}

We consider the total \ion{N}{1} and \ion{H}{1} column
densities to estimate the overall metallicity summed over both components.
In gas that is not severely ionized, \ion{N}{1} is tied to \ion{H}{1} by
charge exchange \citep{Butler:79}, so even if the gas is somewhat affected
by photoionization by UV background light (as one might expect given the
apparent location of the absorber in the CGM), ionization corrections can
be ignored for these two species.  If we assume that ionization
corrections can be ignored for \ion{N}{1}, we obtain
\begin{equation}
\left[ \frac{\rm N}{\rm H} \right] = 
{\rm log} \left( \frac{N(\textsc{N~i})}{N(\textsc{H~i})} \right) - {\rm
  log} \left( \frac{{\rm N}}{{\rm H}} \right) _{\odot}  = -1.0\pm 0.1
\end{equation}

where (N/H)$_{\odot}\:=-7.83\pm0.05$ is the solar nitrogen abundance.

It is well known that in various contexts, nitrogen can be
less abundant than other elements, particularly $\alpha$-capture elements,
compared to solar relative abundances due to its complex nucleosynthetic
origins \citep[e.g.,][and references
therein]{Vila-Costas:93,Pettini:08,Battisti:12}, so this [N/H] measurement
might underrepresent the overall level of enrichment of the sub-DLA.
Therefore, we are motivated to investigate the abundances implied by the
\ion{Si}{2}, \ion{Mg}{2}, and \ion{Si}{3} lines as well.  Unlike the case for 
nitrogen, the corrections for these other species could be important.  For
example, in addition to \ion{Si}{2} being in the same
gas phase as the \ion{H}{1}, some of the \ion{Si}{2} could also arise in more
highly ionized gas that does not contribute significantly to the measured
\ion{H}{1} column density. Indeed, the
detection of \ion{Si}{3} absorption indicates the presence of some
significantly ionized material somewhere in the absorption system.

For these reasons, we have also investigated how ionization corrections could alter the derived
abundances.  In general, the logarithmic gas-phase abundance of ion X$^{i}$ with respect to hydrogen is
\begin{equation}
\left[ \frac{{\rm X}^{i}}{\rm H} \right] = 
{\rm log} \left( \frac{N({\rm X}^{i})}{N(\textsc{H~i})} \right)
- {\rm log} \left( \frac{\rm X}{\rm H} \right) _{\odot} 
+ {\rm log} \left( \frac{f(\textsc{H~i})}{f({\rm X}^{i})} \right),
\label{eqn_abundance}
\end{equation}

where $N$ is the column density, $f$ is the ion fraction, 
[X/H]$_{\odot}$ is the solar abundance of element X with respect to H, 
and $i$ is the ionic species; the
last term in equation~\ref{eqn_abundance} is the ionization correction. We
use the photoionization code \textsc{cloudy} \citep[version
13.03,][]{Ferland:13} to model the ionization of the gas.  Given the \ion{H}{1} column
density of the absorber, some self-shielding could occur and affect the
ionization of the gas, but this can also be explored with \textsc{cloudy}.
Our modeling goal is to test whether we can explain all of the measured
column densities, treating each component individually but assuming that
each component has the same overall metallicity and the same relative
abundances.  We model the absorber as a plane-parallel, constant density
slab photoionized by an external radiation field. Many galaxies are found
within projected distances of a few hundred kiloparsecs of the sightline (see
\S\ref{sect_galaxydata}), so we assume that the ionizing flux
field impinging on the absorber is dominated by flux emerging from these
galaxies.  We approximate the flux field with that of the midplane Milky
Way radiation field modeled by \citet[][]{Fox:05}.\footnote{We also
  examined models in which the gas is ionized by the UV background from
  QSOs \citep[][]{Haardt:96,Haardt:12}.  The result is that our primary
  findings (regarding the abundances of N, Mg, and Si) did not change
  significantly; these various radiation fields have similar shapes near
  the ionization energies of the relevant nitrogen, magnesium, and silicon
  ions, so the models are insensitive to assumptions about the source of
  the ionizing UV flux.}  In photoionization models, the degree of gas
ionization depends on the ionization
parameter $U$ ($\equiv$ ionizing photon density/particle number density),
and the overall abundances depend on the assumed metallicity and relative
abundance patterns (e.g., N/Si); we initially adopted solar relative
abundances, but as we discuss below, we find that some departures from
solar relative abundances are required.  To fit the observed column
densities, we vary the model $N$(\ion{H}{1}), ionization parameter, and
the overall metallicity in each component until the model column densities
agree with the observed values within their $1\sigma$ uncertainties.  We
model each of the absorber components separately to allow for differing
physical conditions, but we require the total \ion{H}{1} column density,
summed over both components, to equal the observed value, 
and we require that both components have the same
metallicity since they are close in velocity and likely arise in regions
that are close in space.

We initially attempted to fit the observed column densities with an overall
metal abundance [M/H] equal to the [N/H] abundance reported above and with
the relative gas-phase abundances of N, Mg, and Si fixed to the solar
relative abundances.  Figure~\ref{fig_cloudy_nodepl} shows the outcome of
this photoionization modeling for component 1 (left panel) and component 2
(right panel).  As expected, the model \ion{N}{1} column density changes
only slightly as $U$ increases; as long as the \ion{N}{1} absorption
arises in the inner region where the gas is mostly neutral, ionization
corrections will be small for \ion{N}{1}.  On the other hand,
$N$(\ion{Si}{3}) increases dramatically with $U$ as the surface of the
gas slab becomes more and more ionized.  We see that
$N$(\ion{Si}{2})/$N$(\ion{N}{1}) also has an appreciable dependence on the
ionization parameter; as $U$ increases, more and more of the Si~\textsc{ii}
absorption originates in the ionized gas on the slab surface where all of
the nitrogen is in the \ion{N}{2} stage (or higher).

For \ion{Si}{2} and \ion{Si}{3}, we see from Figure~\ref{fig_cloudy_nodepl}
that the photoionization models offer a variety of solutions. We first
consider the models for component~1. At low $U$ values such as
the model with log $U = -5.4$ labeled as ``Solution 1'' in
Figure~\ref{fig_cloudy_nodepl}, ionization corrections are negligible, so,
as we expect, the model shows $N$(\ion{Si}{2})/$N$(\ion{N}{1})
$\approx$ (Si/N)$_{\odot}$.  However, at log $U = -5.4$,
we would not expect to detect
any \ion{Si}{3} absorption.  Thus, if the \ion{Si}{2} and
\ion{N}{1} absorption only occurs in the neutral region such as at this
Solution 1, then the \ion{Si}{3} can only be understood by invoking
some separate, more highly ionized \ion{Si}{3} phase that has a 
low-enough $N$(\ion{H}{1}) so that it does not appreciably affect the damping
wings of the Ly$\alpha$ profile.  The \ion{Si}{3} phase also must be
sufficiently ionized so that it contributes little to the \ion{Si}{2}
column but is not so ionized that it leads to detectable \ion{C}{4}
lines (see below).  Although this is physically plausible, it requires some
fine tuning.  Moreover, we also see that
a higher ionization parameter such as ``Solution 2'' (log $U \approx -4.1$)
would produce enough \ion{Si}{3}, so a simpler model of a gas slab with
an ionized surface more fully agrees with the observed data. In this
simpler model, log $U$ cannot exceed $\approx -3.8$ or the model will
produce too much \ion{Si}{2} and \ion{Si}{3}.

The absence of \ion{N}{1} in component 2 suggests that this component is
more highly ionized or has a significantly lower $N$(\ion{H}{1}) than
component 1 (or both), and this is supported by the photoionization
modeling presented in the right panel of Figure~\ref{fig_cloudy_nodepl}: we
see that the silicon ion column densities and the absence of \ion{N}{1}
in component 2 can be easily satisfied with a photoionization model that
has the same gas metallicity as component 1 but a significantly higher
ionization parameter (log $U \approx -3.7$).  We note that a lower
resolution spectrum of PG1543+489 obtained by \citet{Borthakur:13} shows no
evidence of \ion{C}{4}  in this absorption system.  This is consistent
with our photoionization models, which predict that log $N$(\textsc{C~iv})
$\ll$ 12 even at the highest values of $U$ shown in
Figure~\ref{fig_cloudy_nodepl}.  The absence of \ion{C}{4} also suggests
that the absorber does not harbor substantial amounts of $\approx 10^{5}$ K
plasma, but a higher resolution spectrum is required to place stringent
limits on \ion{C}{4}.  A \textit{Far Ultraviolet Spectroscopic Explorer}
spectrum of PG1543+489 does not show significant absorption from
\ion{O}{6} at the absorber redshift, but the data are noisy.

However, our best measurements of the observed \ion{Mg}{2} column densities
(black circles in Figure~\ref{fig_cloudy_nodepl}) are not well fitted in
this model.  In component 1, $N$(\ion{Mg}{2}) has a large uncertainty
because this component is deep and slightly saturated, and we see that the
model is marginally consistent with the measured \ion{Mg}{2} column within
our estimated uncertainties if we choose Solution 1.  However, this
solution does not explain the observed \ion{Si}{3}.  If we choose Solution
2  instead, then the model
\ion{Mg}{2} column is considerably higher (by $\approx$ 0.5 dex) than the
observed column.  Moreover, in component 2 an even more severe discrepancy
between the model and observed $N$(\ion{Mg}{2}) is found at all values of log $U$.

\begin{figure}
\includegraphics[width=8cm, angle=0]{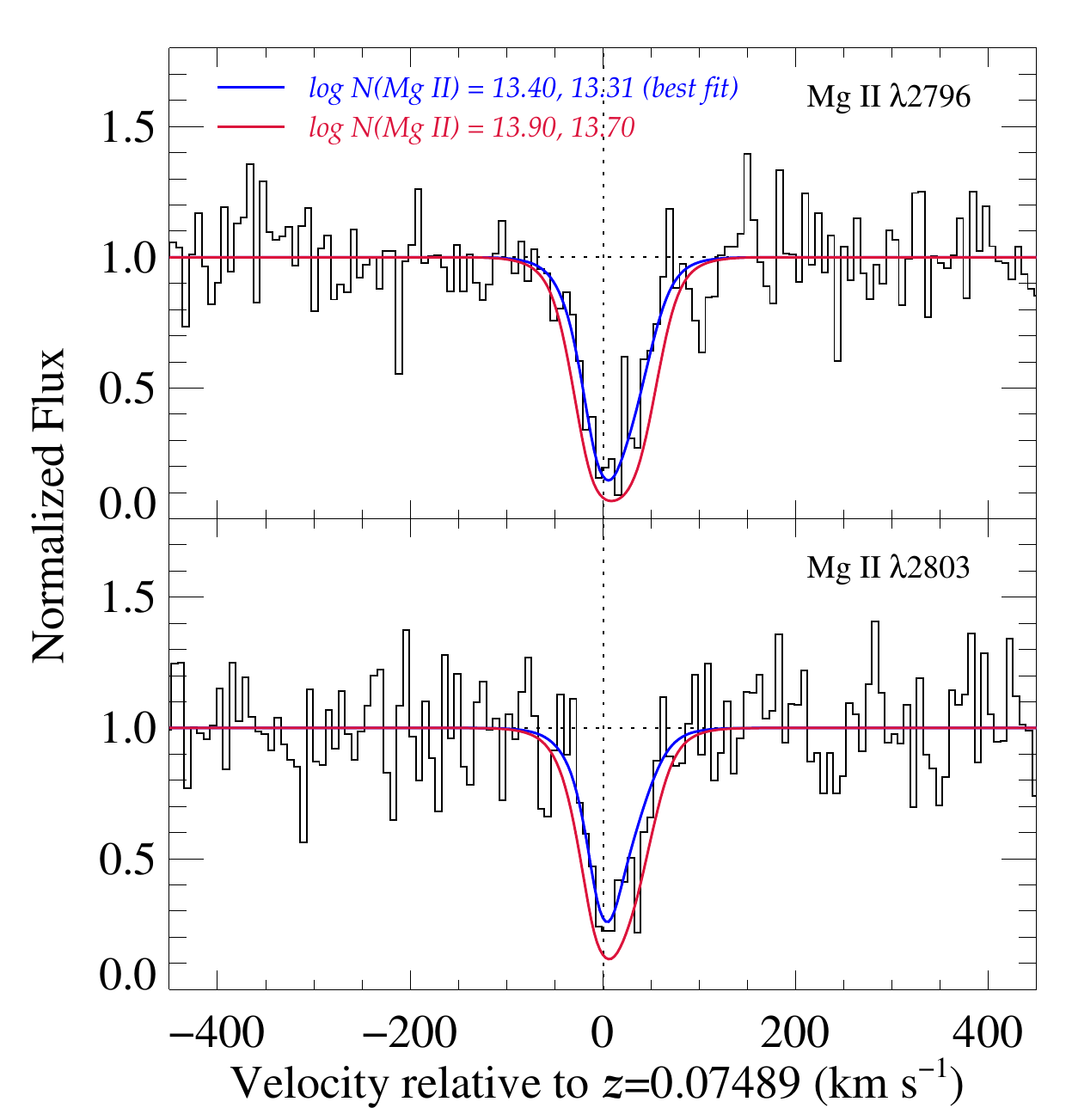}
\caption{Voigt profiles for a two-component model of \ion{Mg}{2}. The blue
  line shows the best-fit profile using the column densities and Doppler
  parameters listed in Table~2, and is the same as that drawn in
  Figure~\ref{fig_spec}; the red line shows the profile assuming $\log
  N$(\ion{Mg}{2})$=13.90$ and 13.70, the approximate values required to
  better match $N$(\ion{Mg}{2}) to the \textsc{cloudy} models in
  Figure~\ref{fig_cloudy_nodepl}. The latter model clearly produces stronger
  absorption than is observed.
\label{fig_badmg2}
}
\end{figure}

A possible explanation for the lower than expected \ion{Mg}{2} column
density might simply be that our profile fitting of the lines
(\S\ref{sect_analysis}) underestimates the true value of
$N$(\ion{Mg}{2}). To demonstrate that this is not likely to be the case, we
show in Figure~\ref{fig_badmg2} profiles (in red) with 
values of $N$(\ion{Mg}{2}) set to better match the \textsc{cloudy}  solutions shown in
Figure~\ref{fig_cloudy_nodepl}, $\log N$(\ion{Mg}{2})$=13.90$ and
13.70. The figure shows that the increased column density produces
theoretical line profiles that no longer fit the data. Of course, the
profiles  are constructed assuming the same $b$
and $v$ values listed in Table~2.  As a test, we refit all of the detected metal
lines listed in Table 2, allowing $b$, $v$ and $N$ to vary, except that
the values of $N$(\ion{Mg}{2}) were fixed to the higher values. The values of $b$,
$v$, and $N$ changed very little as a result, since the model is strongly
constrained by the (higher signal-to-noise ratio) \ion{N}{1} and \ion{Si}{2} lines.
We conclude that the value of $N$(\ion{Mg}{2}) in
Table~2 is robust and does not match the predicted \textsc{cloudy} models
in Figure~\ref{fig_cloudy_nodepl}.

Compared to \textsc{cloudy} models, it appears that our data indicate that
magnesium is underabundant (compared to solar) in this sub-DLA.  At first
glance, one might argue that this could be explained with some amount of
depletion of magnesium and silicon by dust, with the Mg preferentially more
depleted than the Si.  Indeed, the Milky Way ISM dust depletion patterns
reported by \citet[][their Figure~6]{Savage:96} generally show Mg to be
$\approx$0.3 dex more depleted than Si, which is about the right amount to
explain the Mg discrepancy in Figure~\ref{fig_cloudy_nodepl}.  However, to
overcome the debilitating saturation of the \ion{Mg}{2}
$\lambda \lambda$2796.35, 2803.53 doublet that occurs in the Milky Way ISM,
\citet{Savage:96} relied on the much weaker \ion{Mg}{2} $\lambda \lambda$
1239.93, 1240.40 lines.\footnote{The \ion{Mg}{2} $\lambda \lambda$ 1239.93,
  1240.40 doublet is much too weak to be detected in the sub-DLA of \qso .}
After the \citet{Savage:96} depletion patterns were published,
\citet{Fitzpatrick:97}, \citet{Theodosiou:99}, and \citet{Sofia:00}
presented compelling observational and theoretical evidence that the
\ion{Mg}{2} $\lambda \lambda$ 1239.93, 1240.40 oscillator strengths should
be lowered by a factor of 2.  This adjustment erases the preferential
depletion of Mg in the Savage \& Sembach patterns, i.e., after this
adjustment, the dust depletion of Mg and Si is expected to be almost
identical \citep[see also][]{Jenkins:09}. This appears to be true for
dust in the Small Magellanic Cloud (SMC) as well, although the number of
magnesium measurements along SMC sightlines is small \citep{jenkins17}.

We note that we have not used the original \ion{Ca}{2} measurements in the \qso\ sub-DLA
\citep[from][]{Bowen:91} because those measurements have insufficient
spectral resolution to distinguish the two components, but we note that the
lower limit on $N$({\ion{Ca}{2}) reported by Bowen~et~al.\/ is easily
  consistent with the \ion{Ca}{2} column densities expected in these
  photoionization models. In addition, our STIS observations cover the
 \ion{S}{2} 1250 and 1253~\AA\ lines, and since sulfur is an $\alpha$-element,
 these lines potentially provide an important measure of the abundance. Unfortunately,
 the upper limit for $N$(\ion{S}{2}) (Table 2)
 is too insensitive to provide a useful constraint.

\section{\sc Summary and Discussion}
\label{sect_discussion}

As discussed in \S\ref{sect_intro}, we reobserved the \qso\ sub-DLA to seek insight
into its unusual characteristics (e.g., detection of \ion{Ca}{2} at a
relatively large impact parameter in the CGM), but our new data have raised
new questions.  In this section, we first briefly discuss some ideas about
the nature of the absorber, and then provide some remarks about how
these ideas could be tested with future observations.  Before interpreting the data and discussing the nature of this absorption
system, we summarize that it has the following characteristics:

\begin{enumerate}
\item The absorber is located in a galaxy group, with five known galaxies
  within a projected distance of 160 kpc and 21 galaxies within a radius of
  10 Mpc and $\pm$500 km s$^{-1}$.  The closest galaxy is a luminous,
  edge-on spiral galaxy (G1) at $\rho$ = 66 kpc, and moreover, the QSO 
  sightline is close to the major axis of {\bf G1}.  A second luminous spiral
  galaxy (G2) is found at $\rho$ = 119 kpc. G1 and G2 are the brightest
  spirals in the group, and they, and the sub-DLA, are all offset to lower
  velocities than the bulk of the galaxies in the group. Hence, these
  objects may be moving into the group for the first time.  Although we do
  find some ``fuzz'' near the QSO that could, in principle, be related to
  the absorption system, the redshift information (based on the 
  detection of four emission lines in Q1543--K1) 
  indicates that this fuzz is related to the QSO host.

\item A comparison of the velocity of
  the absorption system to the rotation curve of the nearby galaxy G1
  indicates that the absorber could arise in a large gas disk that corotates
with the stellar disk of that galaxy. The absorber velocity is
  lower than an extension of the flat part of the galaxy's rotation curve,
  but that could be due to a Keplerian decrease in the rotation velocity at
  larger radii.  Alternatively, this could indicate that the gas is in a
  rotating gas disk but also has an inflow component.  Several other
  studies have reported similar evidence that galaxies have large rotating
  gas disks with declining rotation velocities at larger distances
  \citep{Steidel:02,Ellison:03,Bowen:16,Ho:17}.  

\item We detect \ion{N}{1}, \ion{Mg}{2}, \ion{Si}{2}, and \ion{Si}{3}
  absorption along with a sub-damped \ion{H}{1} Ly$\alpha$ absorption line,
  and we fitted two-component Voigt profiles to the metal-line data.  The
  \ion{N}{1} and \ion{H}{1} indicate an overall metallicity
  of [N/H] = $-1.0\pm0.1$, or $\simeq 0.1 Z_{\odot}$.
  Photoionization models can explain the observed \ion{Si}{2} and
  \ion{Si}{3} in both components with the same metallicity, but the
  \ion{Mg}{2} indicates that magnesium in the absorber is underabundant
  compared to the model expectations.
 
\end{enumerate}

If we accept the photoionization modeling results for silicon and nitrogen
at face value, then this sub-damped absorber presents some puzzling
characteristics.  The sub-DLA nitrogen abundance is almost an order of
magnitude lower than the metallicity determined from the \ion{H}{2} region
emission lines observed in G1. This is not unprecedented, however: using
COS observations of the QSO PG 1630+377, \citet{Ribaudo:11} have reported a
similar CGM absorber with a metallicity that is a factor of $\approx$30
lower than the metallicity of its closest galactic companion at $\rho$ = 37
kpc.  \citet{Ribaudo:11} suggested that this metallicity discrepancy could
indicate that the low-metallicity absorber is an example of the long-sought
``cold-accretion'' mode \citep{Keres:05} of galaxy gas acquisition.
It is interesting that the sub-DLA shows kinematical signatures of cold accretion,
according to some theoretical works on how cold accretion would
occur \citep{Stewart:11,Stewart:13}.
Indeed, a gradient of only
$-0.014$ dex~kpc$^{-1}$ is needed to match the galaxy's metallicity to that
of the DLA, if indeed the disk could be thought of as extending out to
66~kpc.  This is well within the range of metallicity gradients for galaxy
disks at all redshifts \citep[e.g.][and references\/ therein]{carton18}.


\begin{figure}
\hspace*{-0.5cm}\includegraphics[width=9cm, angle=0]{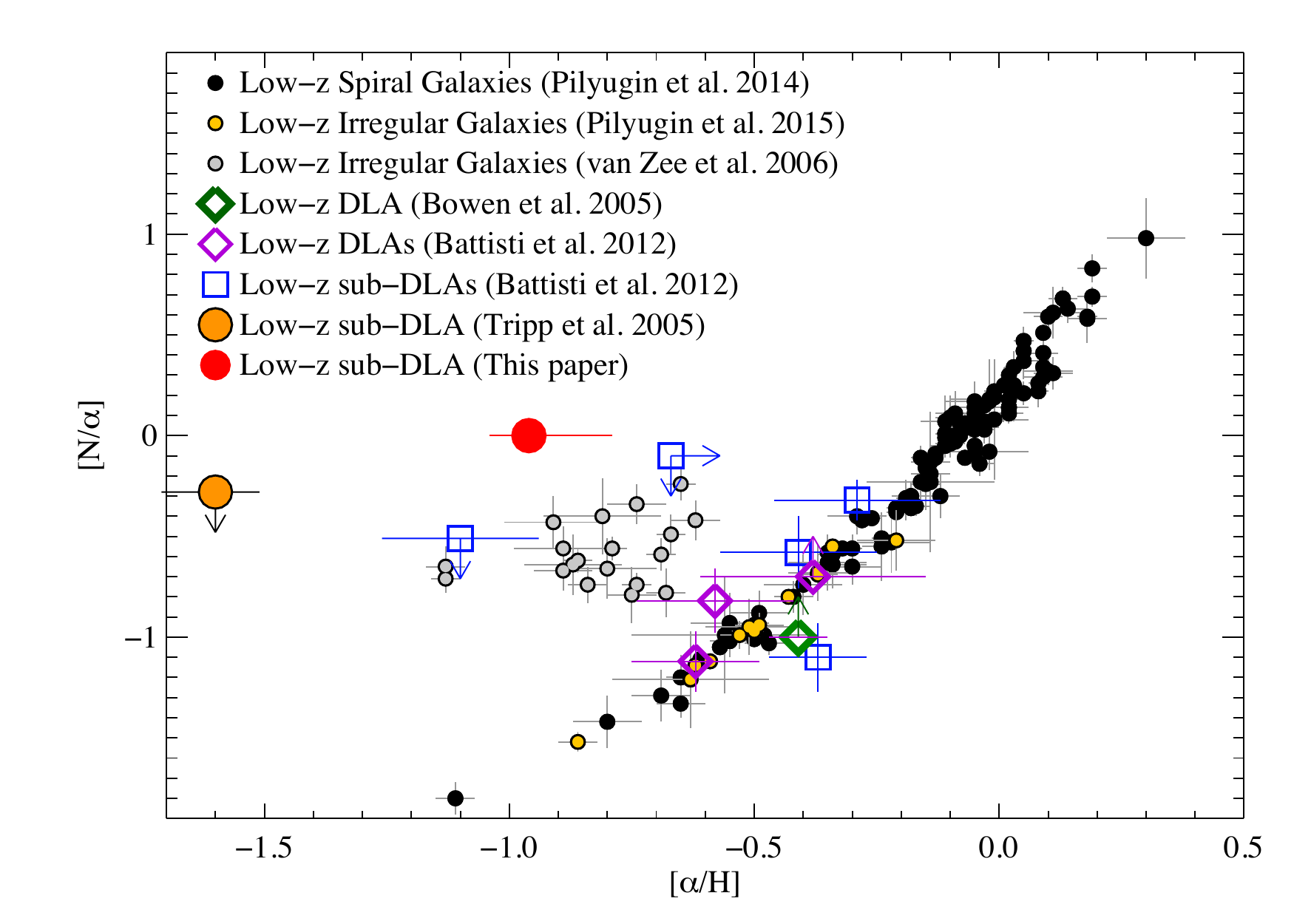}
\caption{Comparison of the unusual abundance pattern in the \qso\ sub-DLA
  (filled red circle) to abundance measurements in galaxies from
  \citet{vanzee:06}, \citet{Pilyugin:14}, and \citet{Pilyugin:15} and to
  abundance measurements in comparably low-redshift sub-DLAs and DLAs from
  \citet{Battisti:12}, \citet{Bowen:05}, and \citet{Tripp:05}.
  The relative abundance of nitrogen compared to an $\alpha$-group
  element is plotted vs.~the overall $\alpha$-group abundance.  Given its
  low [$\alpha$/H] value, it is somewhat surprising to find that the
  [N/$\alpha$] is approximately solar. \label{Fig_1543DLA_vs_gals}}
\end{figure}

Thus, the large discrepancy between the metallicities of the sub-DLA
absorber and galaxy G1 are not necessarily problematic.  The puzzling part
is the apparent consistency of the [N/$\alpha$] relative
abundance\footnote{In this discussion, we are using silicon as a
  representative broader group of ``$\alpha$'' elements.  Silicon is often
  used in the literature as a typical $\alpha$ element.} with the solar
value given the low overall level of metal enrichment of the absorber.  To
show why this is unusual, in Figure~\ref{Fig_1543DLA_vs_gals} we compare
the \qso\ sub-DLA abundance (large red circle) to abundances in
low-redshift galaxies and other low-$z$ DLAs and sub-DLAs (see the 
legend in the upper-left corner).  As we can see from this figure, unlike the \qso\
sub-DLA, most low-$z$ galaxies and DLA/sub-DLA absorption systems have
moderately to significantly subsolar [N/$\alpha$] relative abundances when
the overall metallicity is $\approx$0.1 $Z_{\odot}$. 
{

One hypothesis that could explain this unusual abundance pattern is that we
have found an example of the ``gas recycling'' predicted by some
theoretical studies of the baryon cycle \citep[e.g.,][]{Ford:13}.  In this
scenario, the gas could have been originally ejected from its galaxy of
origin (perhaps G1) with a much higher overall metallicity, which would
also explain the high [N/$\alpha$] relative abundance.  If that ejected gas
subsequently mixed with rather metal-poor/pristine material in the CGM or
the intragroup medium, the mixing might mainly add hydrogen to the
absorber, which could reduce the overall metallicity without changing the
[N/$\alpha$] abundance by much.  Given the observed kinematical correspondence
of the absorber with the nearby galaxy, in this scenario the recycled gas
would have now settled into the gas disk and will eventually feed into the
stellar region of G1.

This hypothesis does not offer an easy explanation of the apparent
underabundance of Mg compared to Si, however.  Magnesium and silicon are
both considered to be $\alpha$-group elements, and in stellar populations,
these elements become overabundant compared to iron as the metallicity
decreases \citep[e.g.,][]{McWilliam:97}, which is a result of the different
nucleosynthetic origins of Mg and Si compared to Fe. The magnitude of the
overabundance of Mg and Si is not always the same in low-metallicity gas,
but unfortunately, the stellar abundance patterns of these two elements go
in the wrong direction to explain the \qso\ sub-DLA abundances: in
low-metallicity objects, Mg is sometimes observed to be more abundant
than Si \citep{McWilliam:97,Prochaska:00}.  It is
possible that the underabundance of Mg is due to preferential depletion of
Mg by dust, but in the most recent studies, Mg and Si are found to be
depleted by dust in very similar amounts \citep[e.g.,][]{Jenkins:09}; there
is no strong evidence that Mg is preferentially depleted (compared to Si)
in other contexts.

 An explanation for the origin of the gas (albeit one
that does not remedy the discrepant Mg/Si ratio) is that the sightline
intercepts debris that arises from interactions between galaxies in
the group, particularly between G1 and G2, which are of similar mass and
could, at least in principle, constitute part of a major interaction.  If
the gas is dominated by material from a galaxy other than G1, it could have
properties that differ from what we found from the emission lines
associated with G1, as described in \S\ref{sect_gal_z}.  Moreover, this gas
could contain dust, which would be responsible for lowering the gas-phase
abundance of Si but not N.  If Si were depleted by $-0.3$ dex, our
observation would be consistent with a true (total) value
[$\alpha$/H]$\: =\: -0.7$, instead of $-1.0$, and [N/$\alpha$]$\:=\: -0.3$.
This condition would still be consistent with our ionization modeling
discussed in \S\ref{sect_cloudy} and would create an outcome in
Figure~\ref{Fig_1543DLA_vs_gals} near the locus of points belonging to the
results of \citet{vanzee:06} for low-$z$ irregular galaxies.  Going
further, a depletion of Si equal to $-0.65$ dex would create the illusion
that [Si/H] (and hence [$\alpha$/H]) equals $-1.0$, while in reality the
value is $-0.35$.  Under this circumstance, [N/$\alpha$]$\: = \: -0.65$,
and this possibility would be consistent with a point on the sequence of
outcomes for low-$z$ spiral and irregular galaxies determined by
\citet{Pilyugin:14,Pilyugin:15}. Both choices would be consistent with our secure
determination that [N/H]$\: = -1.0$.

One of the most
detailed high $N$(\ion{H}{1}) maps of a nearby interacting galaxy group
comes from 21~cm observations of the M81 triplet \citep{yun94,deblok18},
which offers a possible comparison with the group toward \qso. These data
show how interacting galaxies can distribute high $N$(\ion{H}{1}) gas over
many tens of kiloparsecs on the sky, akin to the distances between the sightline
toward \qso\ and G1, G2, and G86.   
Absorption from an interacting group might be expected to be complex,
spanning several hundred \kms , from clouds in different regions of the debris, as
observed toward the line of sight of SN~1993J, which arose in M81 \citep{deboer93,bowen_93j}. We see
no evidence for complex absorption toward \qso, but the environment
intercepted by the sightline to SN~1993J was very different from that
toward \qso, passing through the ISM and CGM of a spiral arm of M81
itself. In addition, the kinematics of any absorbing gas from tidal debris
will depend on the orientation of the interaction on the plane of the sky,
which is unknown for the galaxies toward \qso. Instead, our simple
absorption-line component structure better matches the $5-10$~\kms\
dispersion of the 21~cm emission seen in the outer regions of M81, away
from the centers of the galaxies  \citep[see, e.g. Figure~5
of][]{deblok18}. Similarly, the velocity of the absorption toward \qso\ is
$\sim -50$~\kms\ from the systemic velocity of G1 (Figure~3),
which is consistent with that of 21~cm emission velocities toward various
regions of the M81 triplet.

The existing data provide no definitive answer as to whether tidal debris is the origin
of the absorption toward PG1543+489.  The biggest problem with the hypothesis
is the discrepancy of nearly 1 dex between the absorption-line
abundance and the metallicity of the ISM gas in G1 derived from its
emission lines (\S\ref{sect_galaxydata}). To first order, we would expect
gas ripped from G1 as a result of an interaction to have a similar abundance
to that which we measure in its disk. This discrepancy alone is not inconsistent with assigning interactions
with G1 as the primary origin of the absorption.  For example, the
metallicities of any gas that might contribute to the absorption from 
G2 and G86 are unknown, and although galaxies within a group might be
expected to share star formation histories, data are required to measure
whether their disks and/or CGM have different metallicities. Mixing of gas between two
(or more) group galaxies with different metallicities, and with the IGM in which
the group resides, could result in the deprecation of gas abundances in the
outer regions of tidal debris complexes. Unfortunately, modeling whether
such mixing could account for the lower abundances (or for the discrepant
Mg/Si ratio discussed above) is beyond the scope of this paper.

Future observations of well-detected (but not saturated)
\ion{O}{1} lines would establish the overall $\alpha-$element metallicity
and remove the ambiguity 
caused by possible depletions by dust.
If the oxygen
abundance is higher and therefore [N/$\alpha$] is subsolar, that would
change our interpretation of the abundance patterns.  Likewise,
observations of species that tend to be more severely depleted by dust
(e.g., iron or nickel) would be highly beneficial for sorting out whether
the abundance patterns are due to dust depletion or to the nucleosynthetic
origin and chemical evolution of the gas.  Better constraints on more
highly ionized species would provide valuable constraints on whether
confusion from ionization effects in a multiphase entity are an important
factor. Also, the QSO is bright enough for the optical lines of \ion{Ca}{2}
and \ion{Na}{1} to be recorded at echelle-like resolutions from the
ground, which would better constrain the multicomponent model. 
Higher spatial resolution images of
the galaxies might provide evidence for galaxy interactions through the
detections of disturbed galaxy morphologies and/or tidal tails (and allow a
more detailed study of the fuzz around the QSO, which we have associated
with its host). 

\acknowledgments We appreciate helpful discussions with Celine Peroux and
the use of custom-built 3D visualization software written by Bettie Stobie.
B.\,L.\,F.~acknowledges partial funding  as an Invited Professor at
the Universit\'e de Paris-Sud, Institute d'Astrophysique Spatial d'Orsay in
2016.  B.\,L.\,F~is grateful also for partial funding at the Institute for
Advanced Study in Princeton, and for hospitality at Princeton University in
summer 2015.} Support was
provided for D.\,V.\,B.~through grant GO-08625.01 from the Space Telescope Science
Institute (STScI), which is operated by the Association of Universities for
Research in Astronomy, Inc., under NASA contract NAS5-26555.



\end{document}